\documentclass[runningheads]{llncs}
\usepackage{cite}
\usepackage{amsmath,amssymb,amsfonts}
\usepackage{algorithmic}
\usepackage{graphicx}
\usepackage{hyperref}

\usepackage{subcaption}
\usepackage[ruled,boxed,noalgohanging,lined,linesnumbered]{algorithm2e}
\usepackage{todonotes}
\begin{document}
%\title{Ensemble Feature Selection }
\title{Analysis of ensemble feature selection for correlated high-dimensional RNA-Seq cancer data}
\titlerunning{Ensemble Feature Selection}

\author{Aneta Polewko-Klim\inst{1} \orcidID{0000-0003-1987-7374} \and
Witold R Rudnicki\inst{1,2,3}\orcidID{0000-0002-7928-4944}}
\authorrunning{A. Polewko-Klim \& W. R. Rudnicki}

\institute{Institute of Informatics, University of Białystok, Białystok, Poland \\
\email{anetapol@uwb.edu.pl}
\and
Computational Center, University of Białystok, Białystok, Poland
\and
Interdisciplinary Centre for Mathematical and Computational Modelling,\\ University of Warsaw, Warsaw, Poland}

\maketitle

\begin{abstract}  %250 words
Discovery of diagnostic and prognostic molecular markers is important and actively pursued the research field in cancer research. 
For complex diseases, this process is often performed using Machine Learning.
The current study compares two approaches for the discovery of relevant variables: by application of a single feature selection algorithm, versus by an ensemble of diverse algorithms.
These approaches are used to identify variables that are relevant discerning of four cancer types using RNA-seq profiles from the Cancer Genome Atlas. 
The comparison is carried out in two directions: evaluating the predictive performance of models and monitoring the stability of selected variables.
The most informative features are identified using a four feature selection algorithms, namely U-test, ReliefF, and two variants of the MDFS algorithm. 
Discerning normal and tumor tissues is performed using the Random Forest algorithm. 
The highest stability of the feature set was obtained when U-test was used.
Unfortunately, models built on feature sets obtained from the ensemble of feature selection algorithms were no better than for models developed on feature sets obtained from individual algorithms. 
On the other hand, the feature selectors leading to the best classification results varied between data sets.

\end{abstract}
\keywords{Random forest \and RNA \and Feature selection \and Ensemble learning}
%\linenumber
\setcounter{tocdepth}{1}
%\listoftodos 

%%%%%%%%%%%%%%%%%%%%%%%%%%%%%%%%%%%5

\section{Introduction}

The high-throughput DNA sequencing techniques produce data with tens of thousands probes and each of them could be potentially relevant for diagnostics, prognosis and therapeutics.

Feature selection (FS) techniques are indispensable tools for filtering out irrelevant variables and ranking the relevant ones in molecular biological investigations \cite{vanjimalar_review_2018},\cite{liang_review_2018}. 
The choice of the FS method is very important for further investigation because it greatly limits number of features under scrutiny, allowing to concentrate on most relevant ones. 
On the other hand, FS increases the risk of omitting biological important variables. 

FS methods are typically divided into three major groups, namely filters, wrappers, and embedded \cite{bolon-canedo_ensembles_2019}.
The bias in the filtering FS methods does not correlate with the classification algorithms, hence they generalise better than the other methods. 
Nevertheless, it is well known that individual feature selection algorithms  are not robust with respect to fluctuations in the input data \cite{pes_ensemble_2019}. 
Consequently, application of a single FS algorithm cannot ensure optimal modelling results both in terms of predictive performance and stability. 
This is particularly evident in the integrative analysis of high-dimensional *omics data \cite{meng_dimension_2016}. 

There are numerous FS algorithms that are based on different principles and can generate highly variable results for the same data set. 
The presence of highly correlated features may result in multiple equally optimal set of features and consequently to the instability of FS method.\cite{Kamka2015} 
Such instability reduces the confidence in selected features \cite{pes_ensemble_2019} and their usage as diagnostic or prognostic markers. 
This variability can be to some extent minimised by application of ensemble methods (EFS) that involve combination of different selectors \cite{bolon-canedo_ensembles_2019}.

\subsection{Related work}
The ensemble FS can be broadly assigned to one of two classes: homogeneous (the same base feature selector) and heterogeneous (different feature selectors), \cite{bolon-canedo_ensembles_2019}. 
Regardless of the class, the output of ensemble FS is given either in a form of a final feature set or as a ranking of features. 
Therefore some papers focus on the comparison of different strategies for the ordering of these feature subsets \cite{wang_ensemble_2019}. 
Other researchers are focused on the evaluation of ensembles.
Two quantities of interest are the diversity \cite{seijo2017} and stability of the feature selection process \cite{pes_ensemble_2019,Moulos2013}. 
And though various methods of feature selection have been developed for high-dimensional data, such as high-throughput genomics data, it is still a big challenge to choose the appropriate method for this type of data \cite{liang_review_2018,wenric_using_2018}. 

The stability of FS algorithms for the classification of this type of data has been investigated for instance by Moulos et. al \cite{Moulos2013} and Dessi and Pes \cite{dessi_stability_2015}.
It was shown,  that stability of ensemble feature selection increase only for these FS methods that are intrinsically weak (in term of stability). 
Shahrjooihaghighi et. al \cite{Shahrjooihaghighi2017} proposed an ensemble FS based on the fusion of five feature selection methods (rank product, fold change ratio, ABCR, t-test, and PLSDA) for more effective biomarker discovery. 
The methodology for comparing the outcomes of different FS techniques is presented in \cite{ferreira_comparative_2013}.

Current study is focused on developing and optimisation of a feature selection protocol aiming at identification of biomarkers important for diagnostic of cancers using the results of high-throughput molecular biology experimental methods.  
It is based on ensemble of four diverse feature selection methods and application of classification algorithm that is used to evaluate quality of the set of features. 
 
The protocol was applied to analyse four human cancer tumor types from The Cancer Genome Atlas (TCGA, https://www.cancer.gov/tcga).

In particular the following detailed issues were explored: 
\begin{itemize}
    \item whether application of ensemble of FS methods gives more stable results than individual algorithms;
    \item what is the optimal number of variables for individual algorithms and for ensemble;
    \item whether models built using features returned by ensemble are better than models built using the same number; of variables returned by individual algorithms;
    \item which feature selection algorithm returns best sets of variables? 
\end{itemize}

The main contributions of the current study are following: 
\begin{itemize}
    \item we present a novel perspective of optimization and evaluation of the feature selection for correlated high-dimensional RNA-Seq cancer data;
    \item we compare both the predictive performance of models and the stability of selected feature sets in ensemble feature selection with that of individual FS algorithms; 
    \item we show, that performance of feature selection methods vary between data sets even in for very similar data sets;
    \item we propose to use an ensemble approach as a reference for selecting the method that works best for a particular data set.
\end{itemize}

\section{Materials and Methods}

\subsection{Data}

Four data sets from \href{https://www.cancer.gov/tcga}{The Cancer Genome Atlas} database that contain RNA-sequencing data of tumor-adjacent normal tissues for various typed of cancer were used. \cite{Hammerman_2012,Collisson_2014,Koboldt_2012,Ciriello_2015,Lawrence_2015,zhou2019metascape}
These data set all include a large number of highly correlated and potentially informative features  \cite{peng_large-scale_2015}. 
The preprocesing of data involved standard steps for RNA-Seq data. 
First the log2 transformation was performed. 
Then features with zero and near zero (1$\%$) variance across patients were removed. 
After preprocessing the datasets contain: 
\begin{itemize}
    \item the primary BRCA dataset: 1205 samples (112 normal and 1093 tumor), 20223 variables;
    \item the LUAD dataset: 574 samples (59 normal and 515 tumor), 20172 variables;
    \item the KIRC dataset: 605 samples  (72 normal and 533 tumor), 20222 variables,
    \item the HNSC dataset: 564 samples (44 normal and 520 tumor), 20235 variables.
\end{itemize} 
All data sets are imbalanced, they contain roughly ten times more cancer than normal samples.

\subsection{Methods}

\subsubsection{Filters used for feature selection.}
The procedure outlined above was applied to four filter FS methods, namely, Mann-Whitney U-test \cite{MannWhitney1947}, ReliefF \cite{Relief1992}, \cite{Kononenko1994} and MDFS \cite{mnich2017all,piliszek2019mdfs} in two variants: one-dimensional (MDFS-1D) and two-dimensional (MDFS-2D). 
Since only the ranking of variables is used in the procedure outlined above no corrections of p-value due to multiple testing were necessary. 

\paragraph{U-test} is a robust statistical filter that is routinely used in analysis of *omics data. 
It assigns probability to the hypothesis that two samples corresponding to two decision classes (normal and tumor tissue) are drawn from populations with the same average value. 
The U-test use the p-value to select and rank the features.

\paragraph{MDFS} is a filter, which is based on information theoretical approach, and which can take synergistic effects between variables into account \cite{mnich2017all,piliszek2019mdfs}. 
MDFS also uses  p-values of the test to rank features.
In the current study, 1D and 2D version of MDFS algorithm were used, referred to as MDFS-1D and MDFS-2D, respectively.  

\paragraph{ReliefF} is a filter that computes ranking of importance for variables in the information system, based on the distances in the small-dimensional subspaces of the system \cite{Kononenko1994}. 
Two variants of distance between nearest neighbours, namely, \textit{ReliefFexpRank} and \textit{ReliefFbestK} were tested for the current study. 
Slightly  better results were obtained for the former, hence it was used in all subsequent work. 
This R implementation of algorithm from  \textit{CORElearn} package was used \cite{Core2018}.

\subsubsection{Filter-based feature selection.}
The individual prediction models in k-fold cross-validation for each of four filter FS methods and data sets were constructed. The feature selection process and the learning process from RNA-Seq data set were realized by using the Algorithm \ref{alg_fs}.
%%%%%%%%%%%%%%%%%%%%%%%%%%%%%%%%%%%%%%%%%%%%%%%%%%%%%%%%%%%%%%%%%%%%%%%%%%%%%%%%%
\begin{algorithm}[htbp]
\SetAlgoLined
\DontPrintSemicolon
\SetKwInOut{Input}{input}\SetKwInOut{Output}{output}
\Input{Learning method $l$\\
       Feature selection method $f$\\
       Number of top features N\\
       Data set $D=\{(y_i,x_i)\}_{i=1}^{M}$ with $V = \{v_{1},\ldots,v_{p}\}$ features \\
       and with M instances, randomly partitioned into about \\
       equally-sized folds $S_{j}$}\
\Output{Ranked feature sets $F_{j}$, $j=1,\ldots, k$\\
       %$n$ predictions for the response variable $\P_i$, $i=1,\ldots, N$\\
       Performance estimation metric E\;
}
\ForEach{$S_j$}{
  Define the training set $D_{\setminus j}(V) \leftarrow D(V)\setminus S_j$(V)\;
  Perform feature selection on the training set $R_{j} \leftarrow f(D_{\setminus j}(V))$\;
  Remove highly correlated features with ranked list $R_{j}$ of features\;
  Collect the N highest ranked feature set $F_{j} = \{v_{1},\ldots,v_{n}\}$ with $R_{j}$ \; 
  Build the model on the training set $L_j\leftarrow l(D_{\setminus j}(F_{j}))$\;
  Performance estimation: use the trained model $L_j$ on a test set $S_j$\;
  $E \leftarrow \frac{1}{k}\Sigma E_{j}$\;
 }
 \caption{${\bf FS}(l,f,N, D=\{S_{1},\ldots,S_{k}\})$ the filter feature selection algorithm with Random forest classifier }
 \label{alg_fs}
\end{algorithm}

%%%%%%%%%%%%%%%%%%%%%%%%%%%%%%%%%%%%%%%%%%
\begin{algorithm}[htbp]
\SetAlgoLined
\DontPrintSemicolon
\SetKwInOut{Input}{input}\SetKwInOut{Output}{output}
\Input{Learning method $l$\\
       The 4 $\times$ k sets of top-N uncorrelated features with 4 filters $W_i$ \\
       $F_{i,j}, i=1,\ldots, 4, j=1,\ldots, k$ with top-N features\\
       Data set $D=\{(y_m,x_m)\}_{m=1}^{M}$ described with features $F_{i,j}$\\
       and with M instances, randomly partitioned into about \\
       equally-sized folds $S_{n}$}\
\Output{Collected feature sets $C_{p}$, $p=1,\ldots, k$\\
       %$n$ predictions for the response variable $\P_j$, $j=1,\ldots, N$\\
       Performance estimation metric $E_{p}$\;
}
\ForEach{$S_j$}{
  Collect the union of feature set $C_{p} = F_{1j} \cup F_{2j} \cup F_{3j} \cup F_{4j}$  \; 
  Define the training set $D_{\setminus n}(C_{p}) \leftarrow D(C_{p})\setminus S_n(C_{p})$\; 
  Build the model on the training set $L_{n,p}\leftarrow l(D_{\setminus n}(C_{p}))$\;
  Performance estimation:  $E_{n,p} \leftarrow L_{n,p}(S_n(C_{p}))$\;
  $E_{p} \leftarrow \frac{1}{k}\Sigma E_{n,p}$\;
 }
 \caption{${\bf EFS}(l,W = \{F_{11},\ldots,F_{4k}\},D=\{S_{1},\ldots,S_{k}\})$ the ensemble feature selection algorithm with Random forest classifier}
 \label{alg_efs}
\end{algorithm}

%%%%%%%%%%%%%%%%%%%%%%%%%%%%%%%%%%%%%%%%%%%%%%%%%%%%%%%%%%%%%%%%

This algorithm outlined above was repeated for several values of $N$ and it was repeated multiple times, to minimize the effects of random fluctuations. 
The stability of feature selection was measured by comparing feature sets obtained in multiple runs of the procedure. 
%%%%%%%%%%%%%%%%%%%%%%%%%%%%%%%%%%%%%%%%%%%%%%%%%%%%%%%%%%%%%%%%%%%%%%%%%%%%%%%

\subsubsection{Ensemble feature selection.}
 
The ensemble set of N-top relevant variables was constructed by a union of top-N variables from each filter FS methods, as it is shown in Algorithm \ref{alg_efs}.
The size of the set may vary between $N$ and $4N$ depending on the similarity of the sets returned by individual FS algorithms. All comparisons between feature sets obtained from ensemble and features sets obtained by individual filters were performed on sets with comparable numbers of total variables. 
For example if union of four top-$5$ sets resulted in a set with 20 variables it was subsequently compared with other sets containing 20 variables. 

All applied filters provide their own ranking of the features.
The U-test and MDFS algorithms rank features by their statistical significance and, the ReliefF algorithm by their performance in classification. 
Then the joint set of most important variables is created as a union of top-N sets from individual rankings. 
The ranking within the combined set was not necessary and it was never performed.

Algorithm \ref{alg_efs} was repeated several times for different values of $N$  as in the case Algorithm \ref{alg_fs}. The stability of feature selection was also estimated.

\subsubsection{Classification}
\begin{figure}[ht!]\centering 
\includegraphics[width=1.0\textwidth]{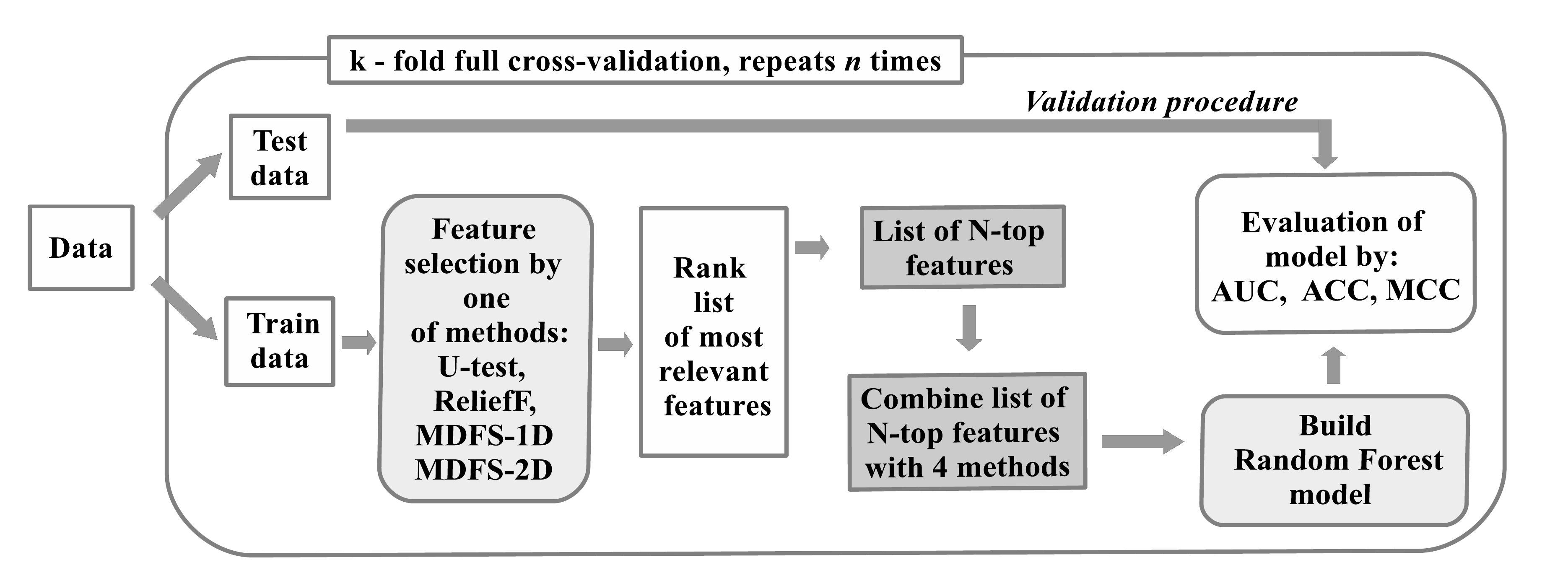} 
\caption{Pipeline of the ensemble FS method. See notation in text.}
\label{fig1}
\end{figure}

The quality of the feature-set was evaluated by building a machine learning model using selected features and measuring its quality. 
To this end the Random Forest \cite{Breiman2001} algorithm was used. 
It has been shown that Random Forest is generally reliable algorithm, that works well out-of the box, rarely fails and in most cases returns results that are very close to best achievable for given problem \cite{JMLR:v15:delgado14a}.
The quality of model was evaluated using area under ROC curve (AUC). 
This measure is independent of the balance of classes in the data and does not need any fitting. 
The scheme of ensemble feature selection and supervised classification is presented in Figure~\ref{fig1}. 

\subsubsection{Measuring stability of feature selection}
The total stability of filter FS method is measured as the average of the pairwise similarity for all pairs of the most informative feature subsets ($s_{i}$,$s_{j}$) from \textit{n} runs of a model in full \textit{k}-fold cross-validation. 
To this end the Lustgarten's stability measure (ASM) \cite{Lustgarten2009}, which can be applied to sets of unequal sizes, was used. 
It is described by the formula:
\begin{equation}
ASM = \frac{2}{c(c-1)}\sum^{c-1}_{i=1}\sum^{c}_{j=i+1}
\left(
\frac{\left|s_{i}\cap s_{j}\right|-\left|s_{i}\right|*\left|s_{j}\right|/m}
{min(\left|s_{i}\right|,\left|s_{j}\right|)-max(0,\left|s_{i}\right|+\left|s_{j}\right|-m)}
\right)
\end{equation}
where: \textit{m} is total feature number of dataset and \textit{c} = \textit{n}*\textit{k}. \\

\subsubsection{Optimization of feature selection.}

In the first step four threshold levels for defining highly correlated variables were examined to establish threshold leading to best results of classification. 
Four thresholds levels were tested: $|r|=\{0.7,0.75,0.8,0.9\}$
The subsequent analyses were performed for the optimal threshold level. 

The following analyses were performed for each individual FS filter and for ensemble FS filters: 
\begin{itemize}
\item how many uncorrelated variables should be included in the model to obtain best classification;
\item how stable is stability measure for top-N feature subsets;
\item whether adding the highly correlated variables to top-N variables influences  predictive power.
\end{itemize}

Entire modelling protocol, including bot feature selection and model building step was performed within $k = 5$ fold cross-validation and was repeated $n = 30$ times, independently for each FS method and data set. 
Within each cross-validation iteration feature selection algorithm was performed once and then models were trained for  all feature set sizes  $N = \{5,10,\ldots ,200\}$.\\

Analysis was performed using the R (version 3.5) \cite{Rcore} and R/Bioconductor packages. \cite{Bioconductor2015}

\begin{figure}[tb]
\centering
  \begin{subfigure}[b]{.24\linewidth}
	  \centering
		\caption{U-test}
		I
    \includegraphics[width=0.84\textwidth,trim ={ 1cm 1cm 1.0cm 2.0cm },clip]{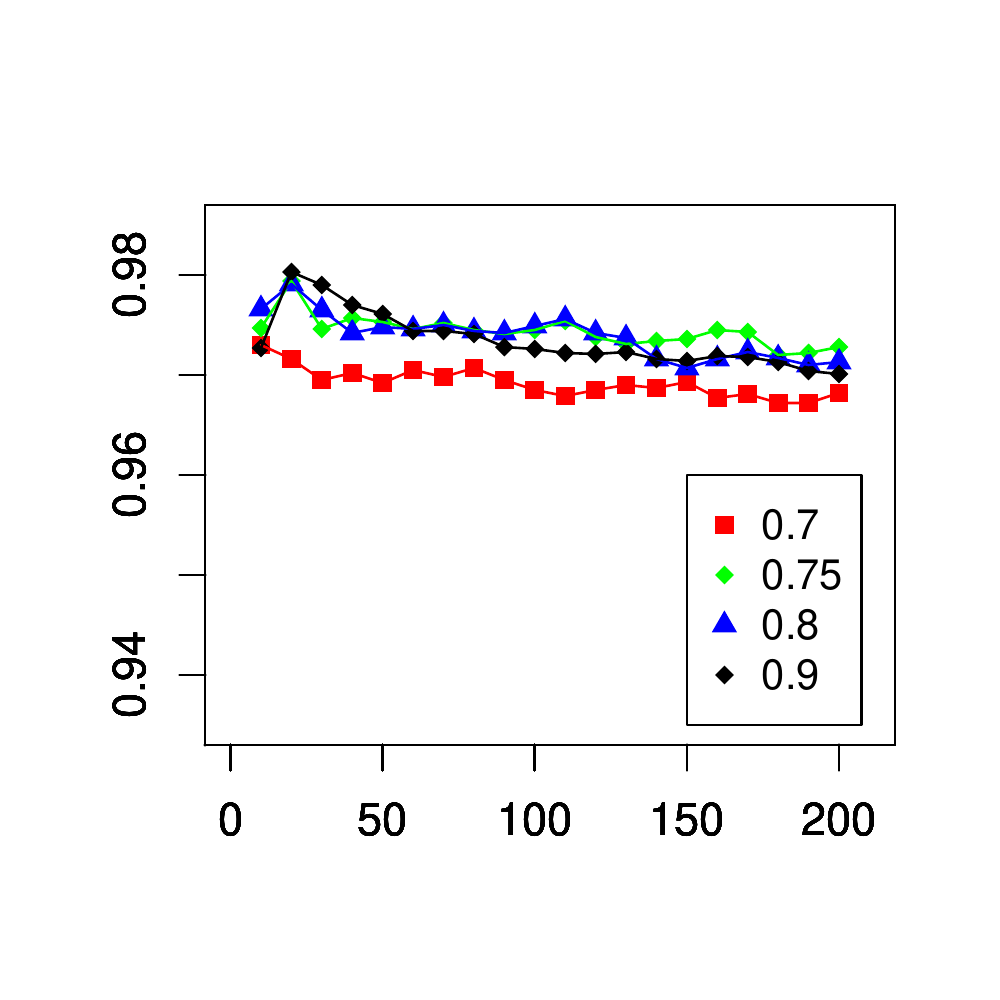} 
  \end{subfigure}%  
  \begin{subfigure}[b]{.24\linewidth}
    \centering
		\caption{MDFS-1D}
    \includegraphics[width=0.84\textwidth,trim ={ 1cm 1cm 1.0cm 2.0cm },clip]{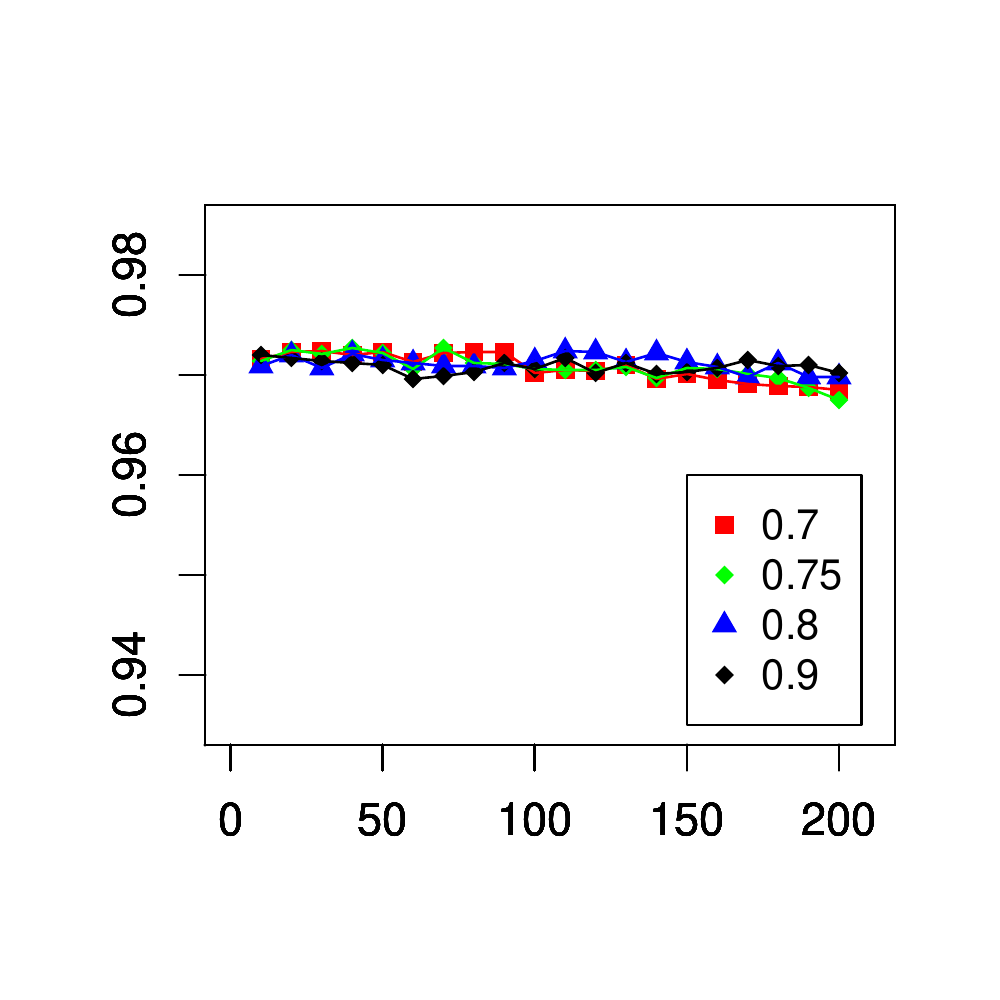} 
  \end{subfigure}
  \begin{subfigure}[b]{.24\linewidth}
    \centering
		\caption{MDFS-2D}
    \includegraphics[width=0.84\textwidth,trim ={ 1cm 1cm 1.0cm 2.0cm },clip]{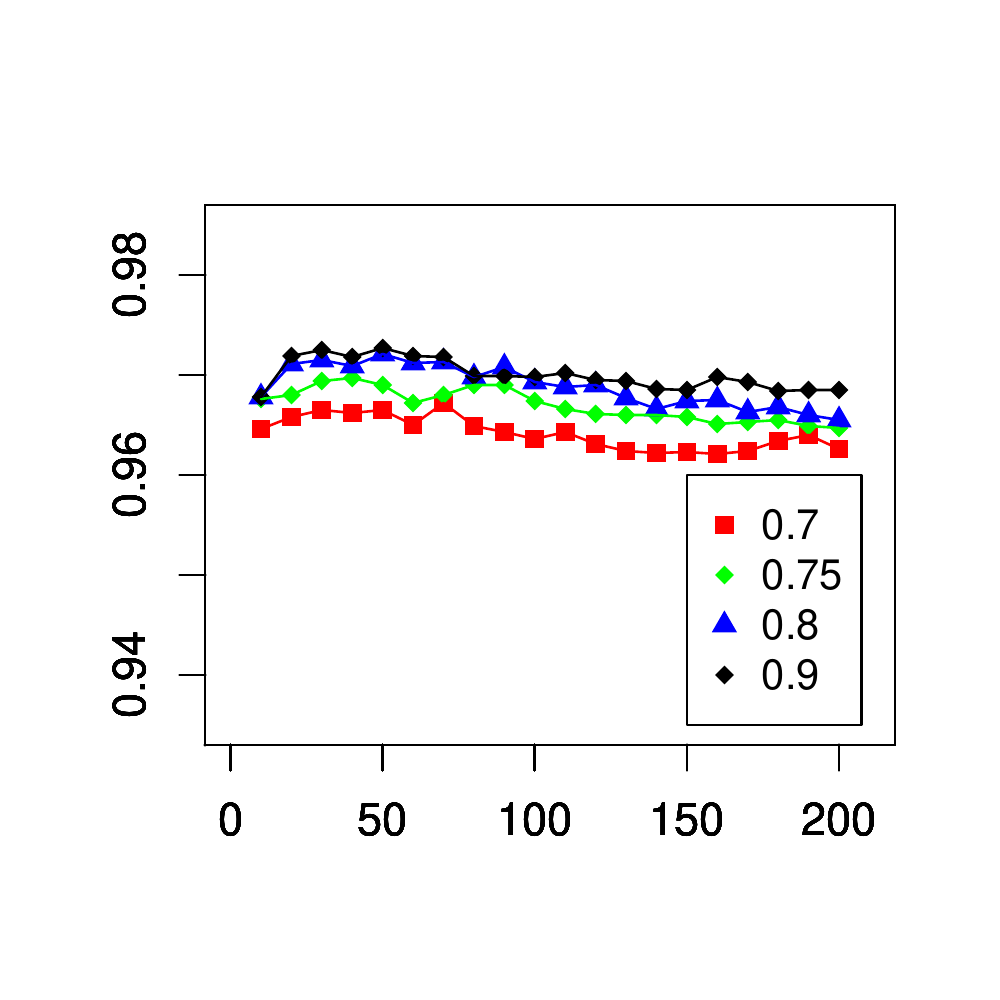}
  \end{subfigure}
  \begin{subfigure}[b]{.24\linewidth}
    \centering
		\caption{ReliefF}
   \includegraphics[width=0.84\textwidth,trim ={ 1cm 1cm 1.0cm 2.0cm },clip]{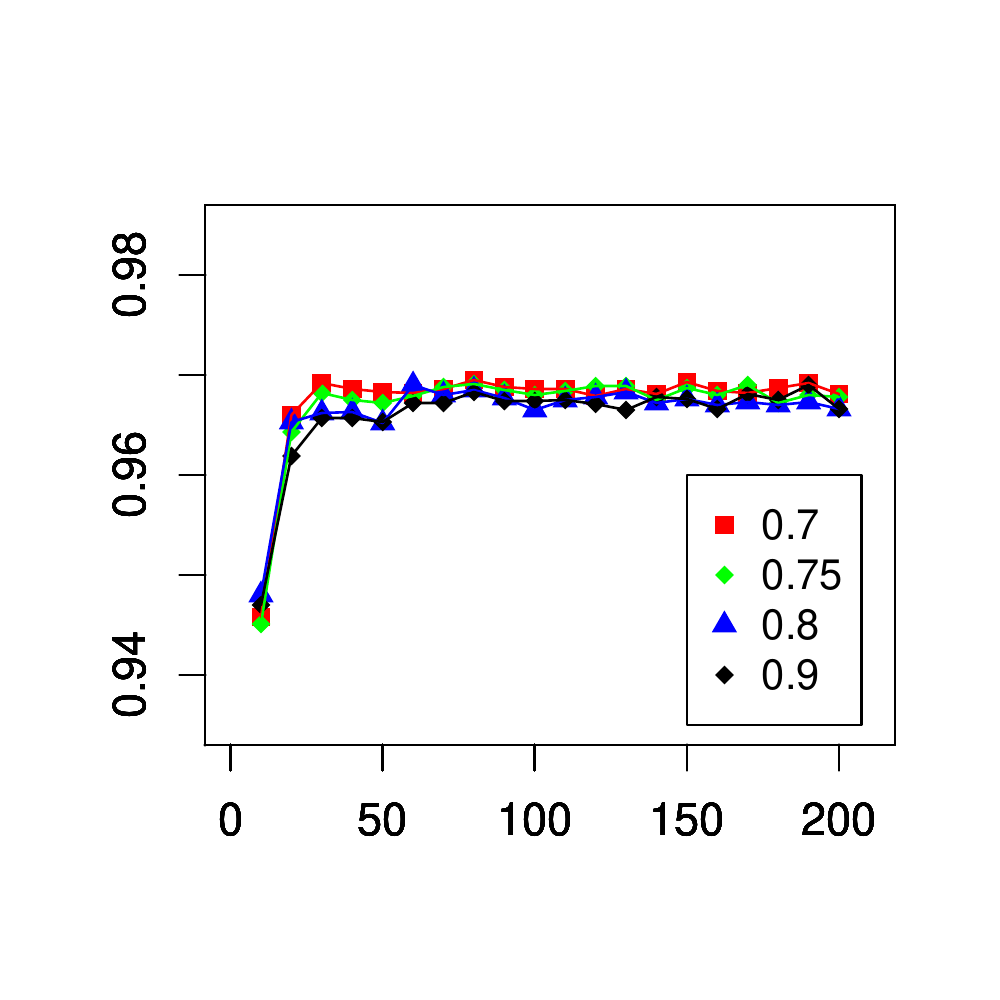}
  \end{subfigure}\\
	  \begin{subfigure}[b]{.24\linewidth}
	  \centering
		II
    \includegraphics[width=0.84\textwidth,trim ={ 1cm 1cm 1.0cm 2.0cm },clip]{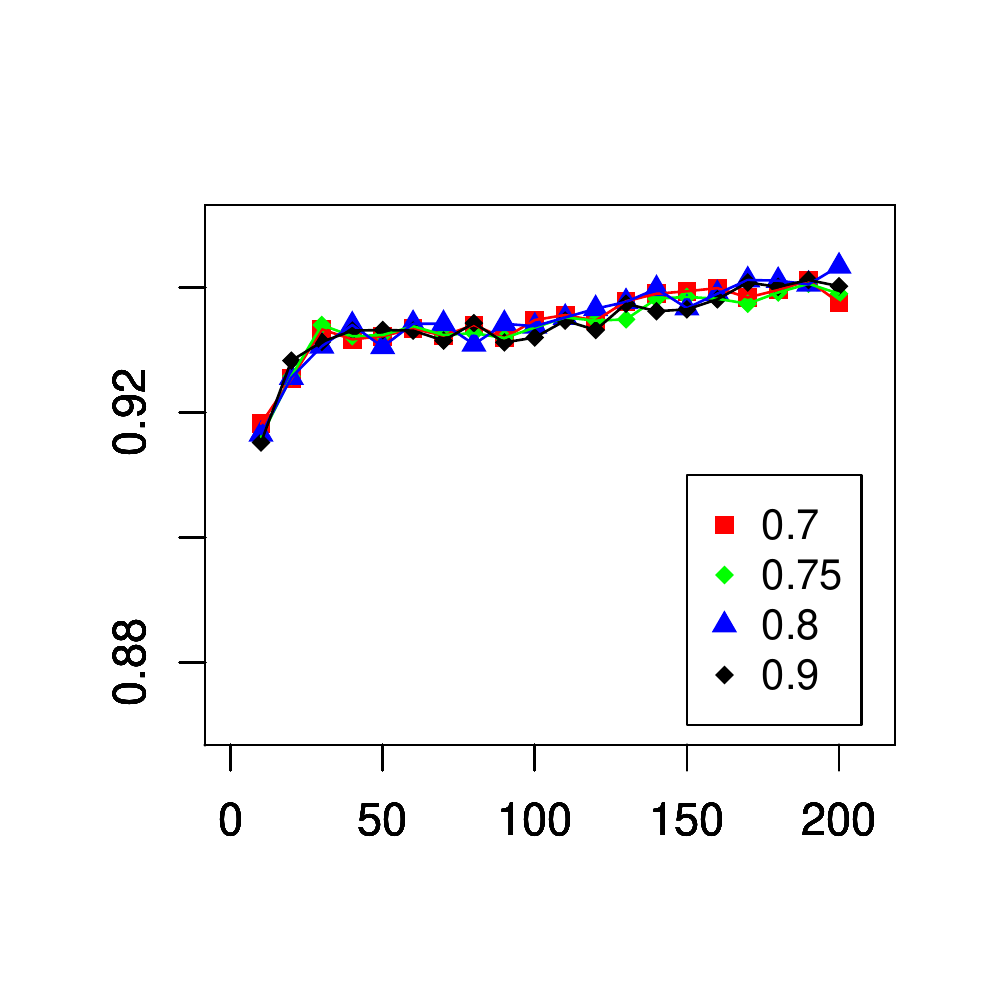} 
  \end{subfigure}%  
  \begin{subfigure}[b]{.24\linewidth}
    \centering
    \includegraphics[width=0.84\textwidth,trim ={ 1cm 1cm 1.0cm 2.0cm },clip]{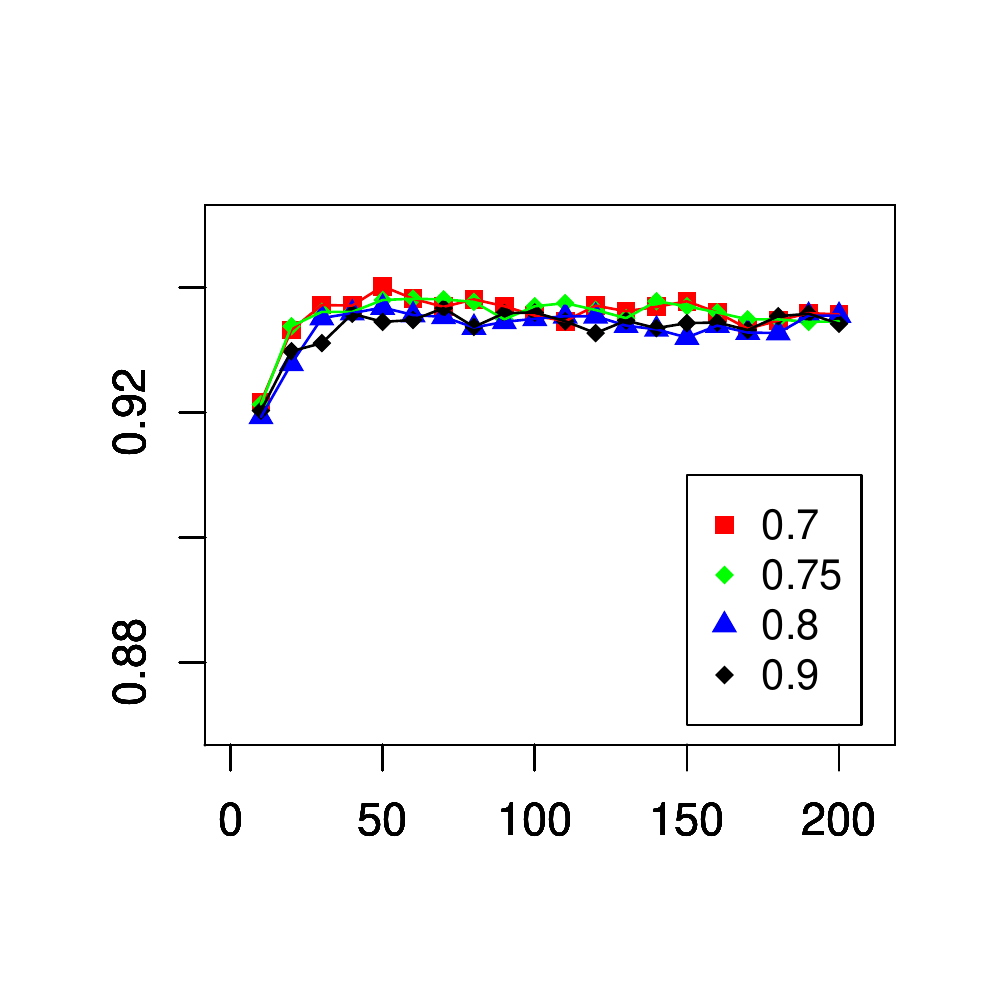} 
  \end{subfigure}
  \begin{subfigure}[b]{.24\linewidth}
    \centering
    \includegraphics[width=0.84\textwidth,trim ={ 1cm 1cm 1.0cm 2.0cm },clip]{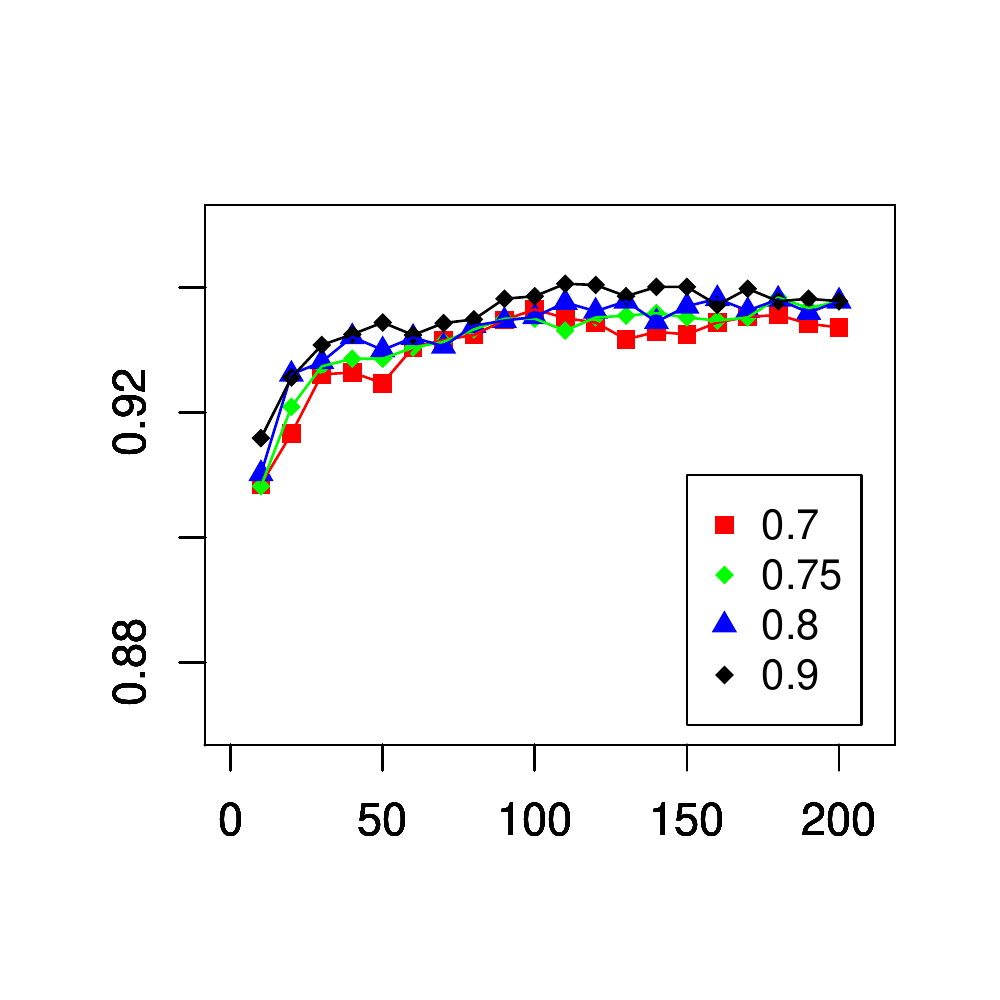}
  \end{subfigure}
  \begin{subfigure}[b]{.24\linewidth}
    \centering
   \includegraphics[width=0.84\textwidth,trim ={ 1cm 1cm 1.0cm 2.0cm },clip]{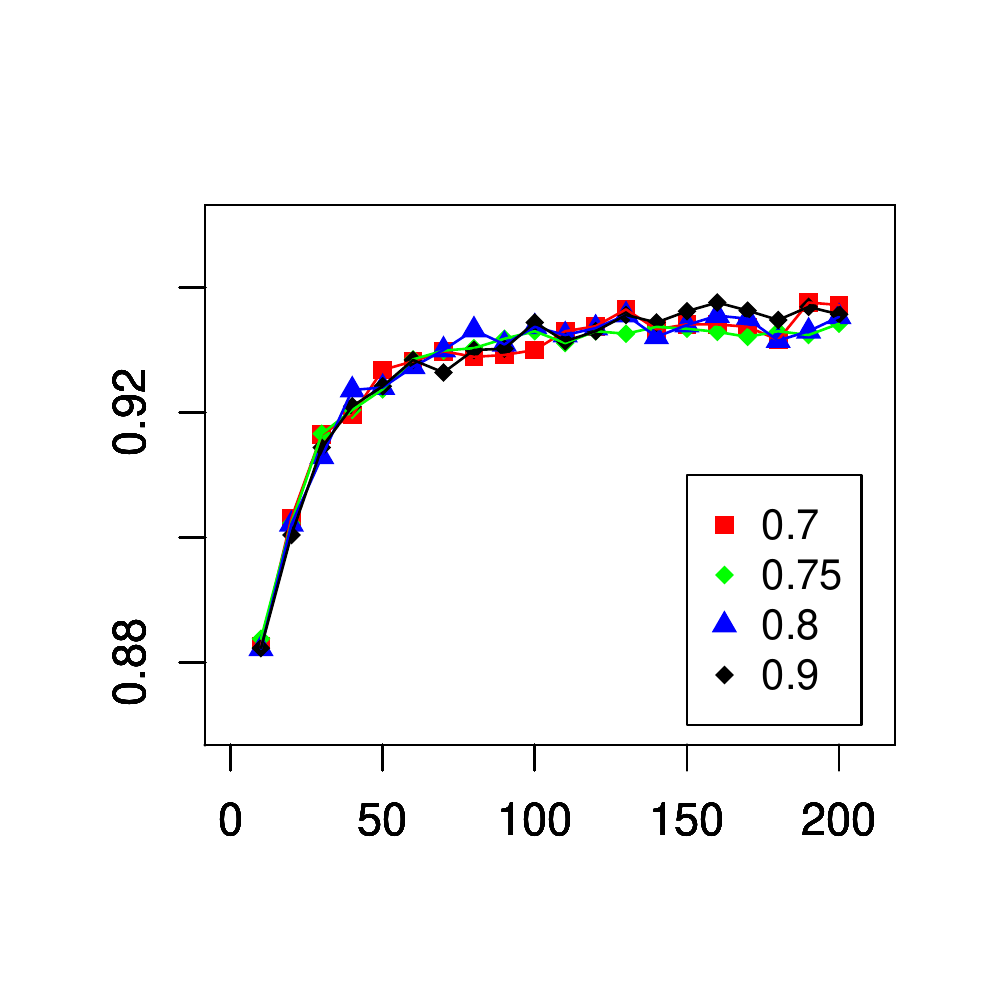}
  \end{subfigure}\\
	\begin{subfigure}[b]{.24\linewidth}
	\centering
	III
  \includegraphics[width=0.84\textwidth,trim ={ 1cm 1cm 1.0cm 2.0cm },clip]{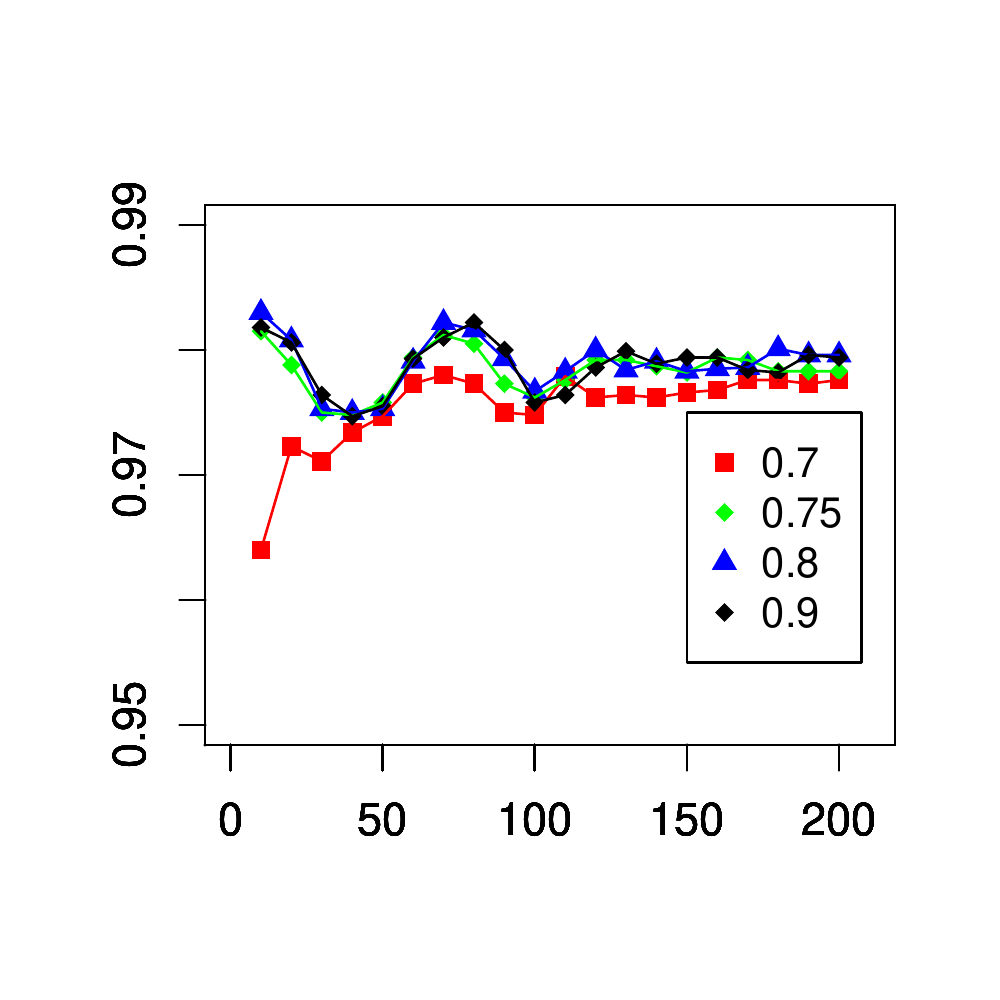} 
  \end{subfigure}%  
  \begin{subfigure}[b]{.24\linewidth}
    \centering
    \includegraphics[width=0.84\textwidth,trim ={ 1cm 1cm 1.0cm 2.0cm },clip]{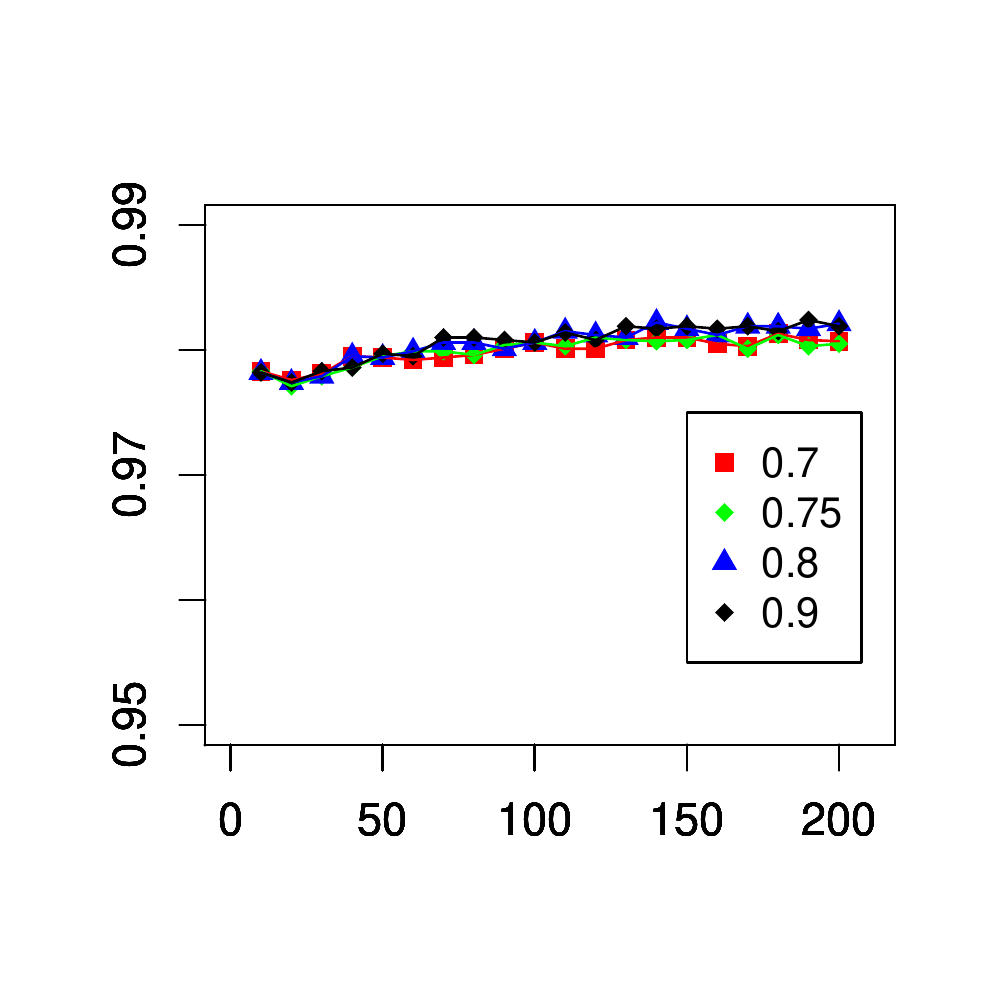} 
  \end{subfigure}
  \begin{subfigure}[b]{.24\linewidth}
    \centering
    \includegraphics[width=0.84\textwidth,trim ={ 1cm 1cm 1.0cm 2.0cm },clip]{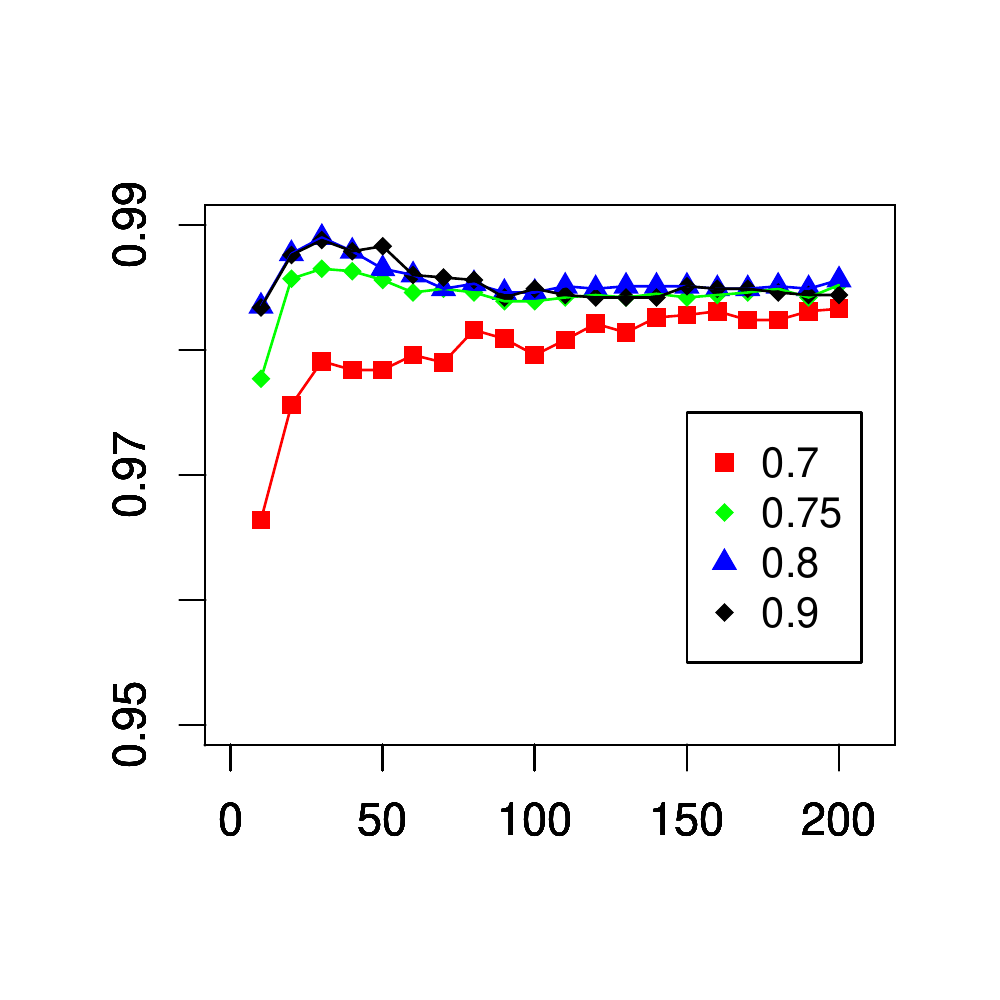}
  \end{subfigure}
  \begin{subfigure}[b]{.24\linewidth}
    \centering
   \includegraphics[width=0.84\textwidth,trim ={ 1cm 1cm 1.0cm 2.0cm },clip]{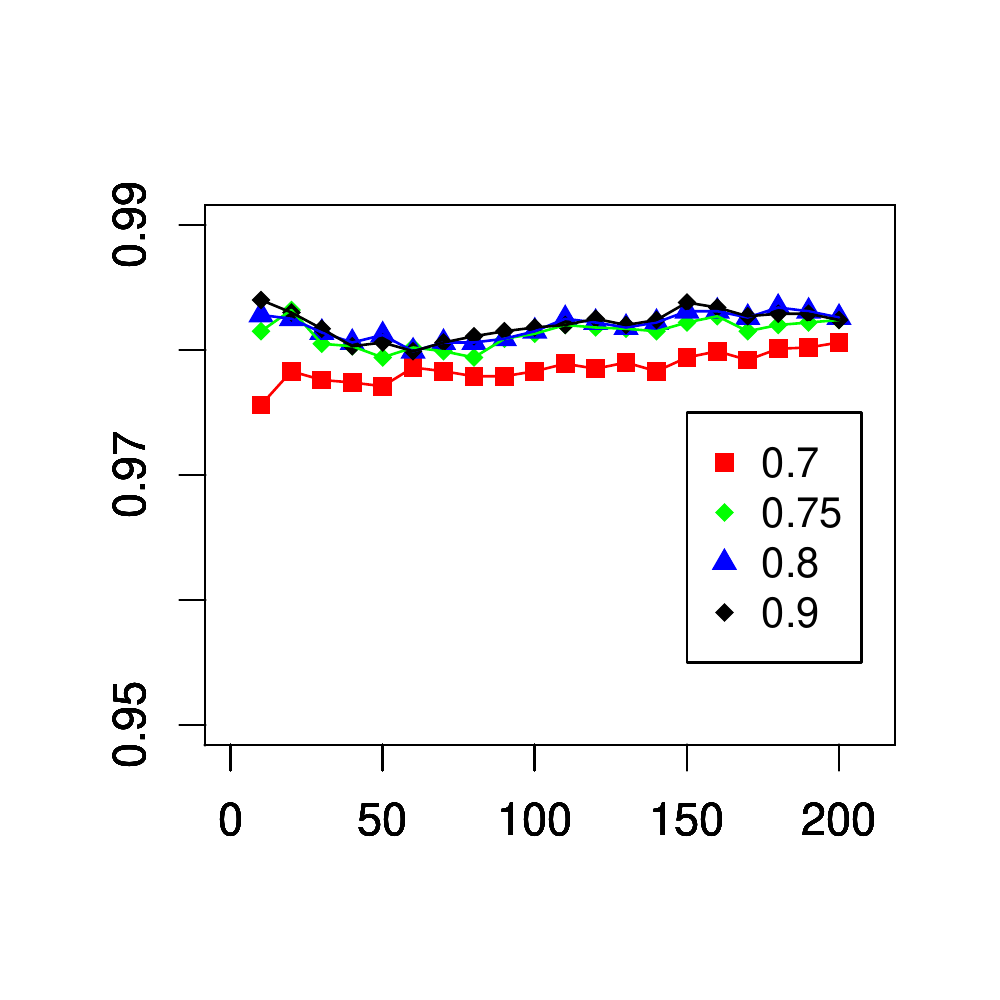}
  \end{subfigure}\\
		  \begin{subfigure}[b]{.24\linewidth}
	  \centering
		IV
    \includegraphics[width=0.84\textwidth,trim ={ 1cm 1cm 1.0cm 2.0cm },clip]{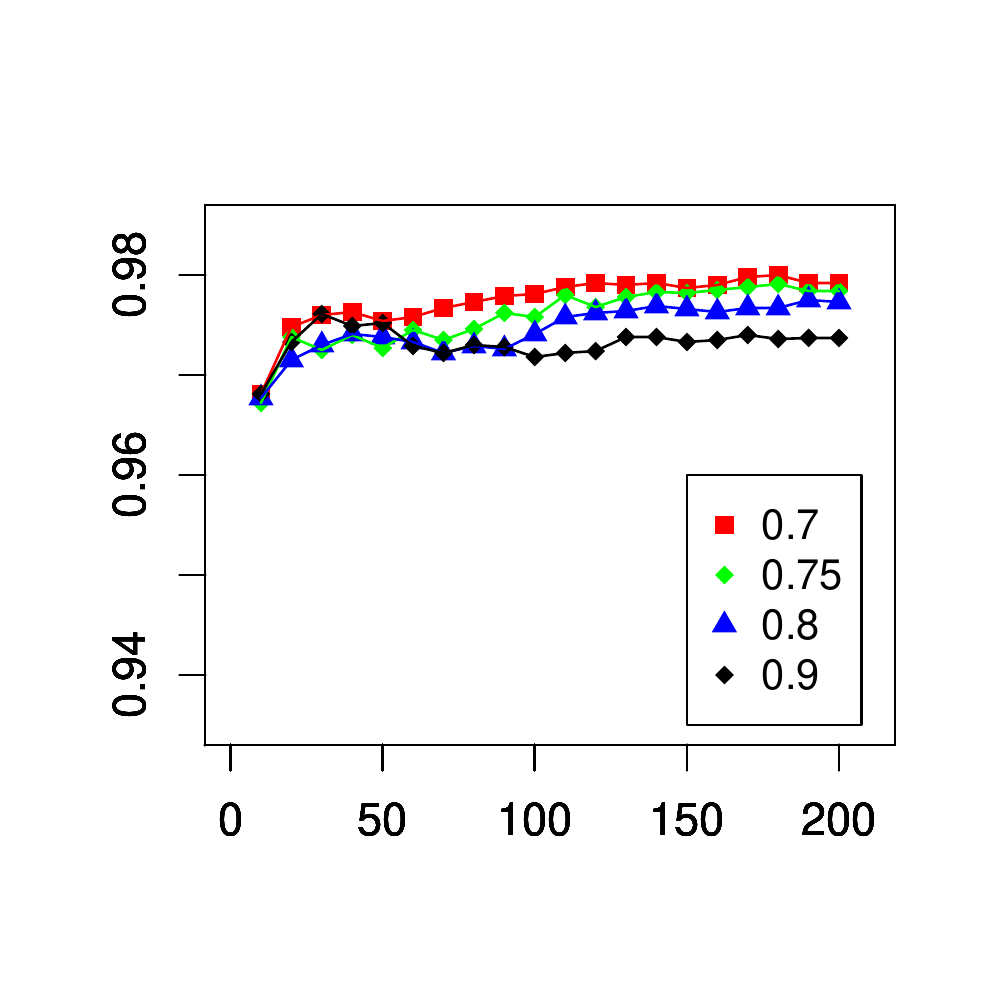} 
  \end{subfigure}%  
  \begin{subfigure}[b]{.24\linewidth}
    \centering
    \includegraphics[width=0.84\textwidth,trim ={ 1cm 1cm 1.0cm 2.0cm },clip]{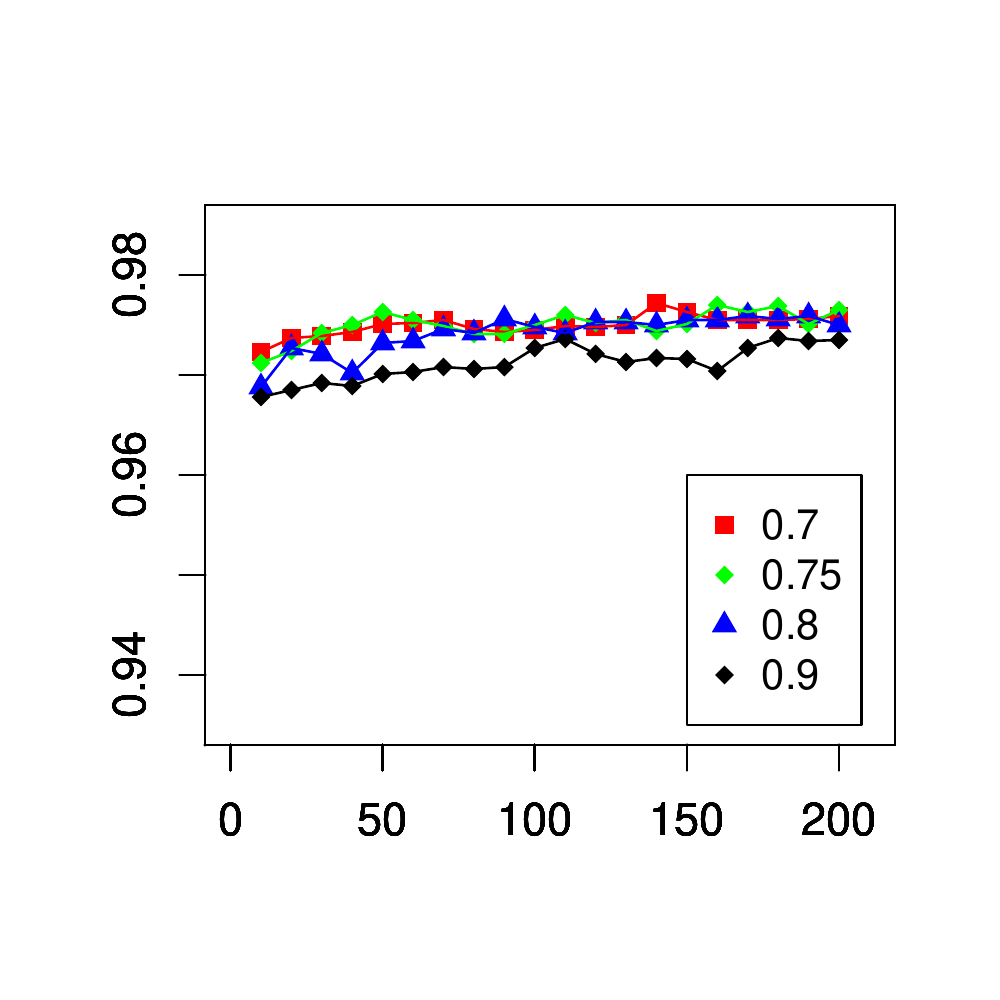} 
  \end{subfigure}
  \begin{subfigure}[b]{.24\linewidth}
    \centering
    \includegraphics[width=0.84\textwidth,trim ={ 1cm 1cm 1.0cm 2.0cm },clip]{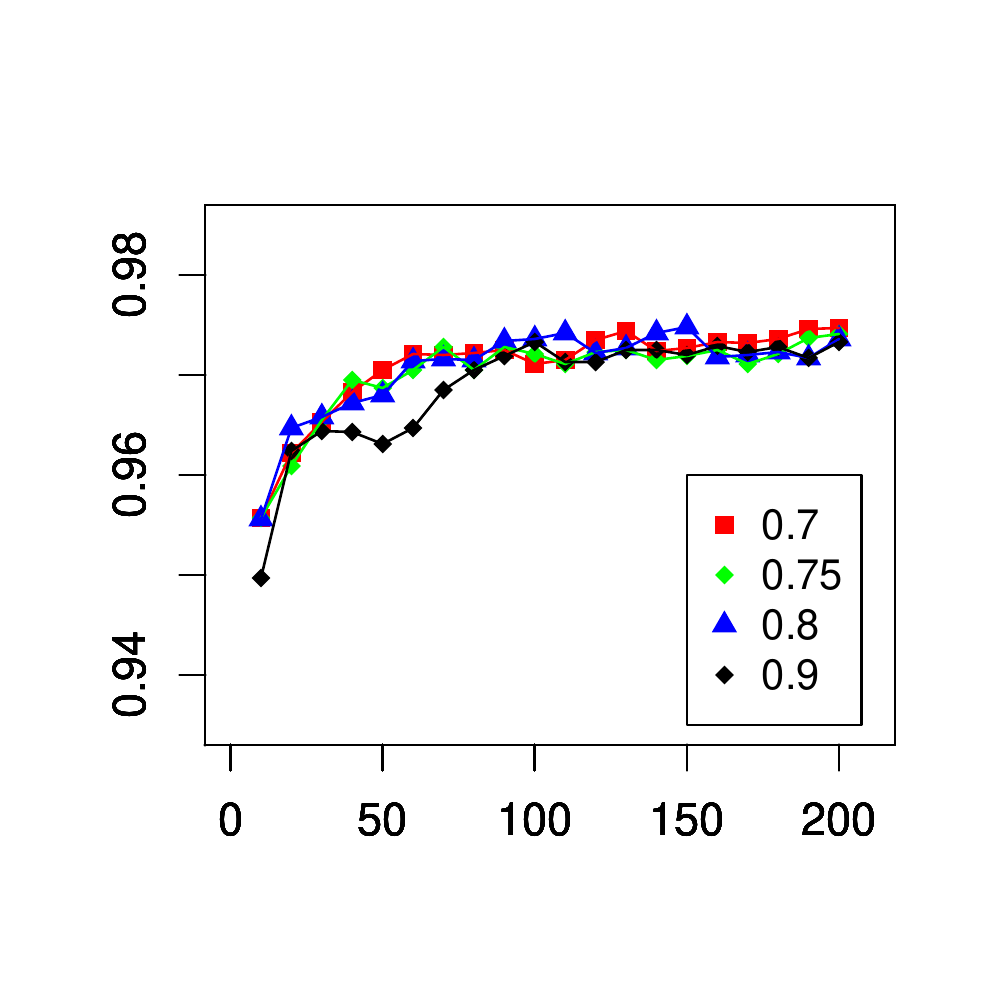}
  \end{subfigure}
  \begin{subfigure}[b]{.24\linewidth}
    \centering
   \includegraphics[width=0.84\textwidth,trim ={ 1cm 1cm 1.0cm 2.0cm },clip]{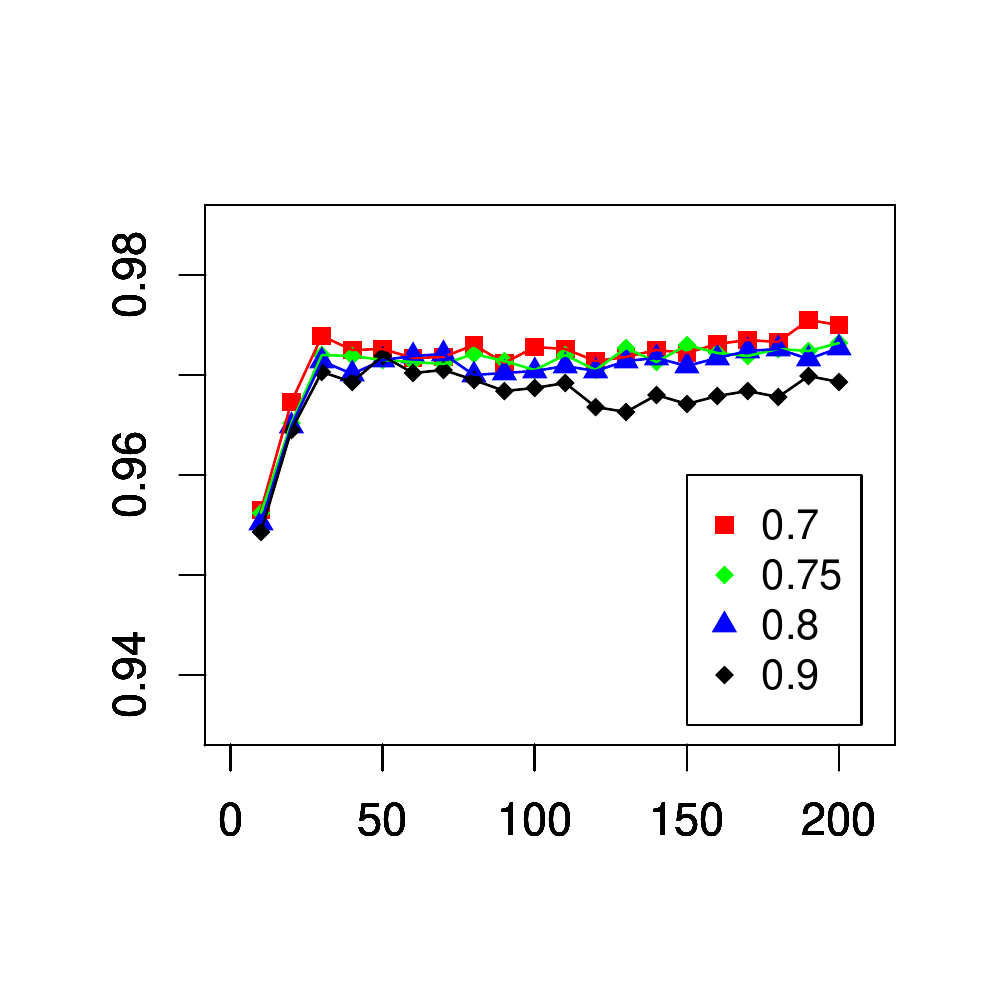}
  \end{subfigure}
\caption{AUC vs N-top biomarkers for different value of Spearman's rank correlation coefficient. Results for BRCA, HNSC, KIRC and LUAD data sets are displayed in rows I, II, III, and IV, respectively.}
\label{fig2}
\end{figure}

%%%%%%%%%%%%%%%%%%%%%%%%%%%%%%%%%%%%%%%%%%%%%%%%%%%%%%%%%%%%%%%%%%%%%%%%%%%%%%%%%%%%%%%%%%%%%%%%%%%
\section{Results and Discussion}

\subsection{Model accuracy}
In the first step, the impact of correlation between informative features on the predictive power of RF model was examined. 
The results of this analysis are displayed in Figure~\ref{fig2}. 
It can be seen that squares corresponding to correlation threshold 0.7 in many cases fall bellow other lines on the AUC plots.  
Therefore threshold for removal of highly correlated variables was set at Spearman's correlation coeffcient $r$ higher than 0.75. 
This value of coefficient is applied in the subsequent analysis. 
One may note, that MDFS-1D filter is the most robust with respect to change in the feature level correlation among the applied FS methods.\\

\begin{figure}[tb]\flushleft 
\centering
  \begin{subfigure}[b]{.45\linewidth}
	  \centering
		\caption{BRCA}
	\includegraphics[width=1.15\textwidth,trim ={  .1cm .1cm 0.1cm 1.8cm},clip]{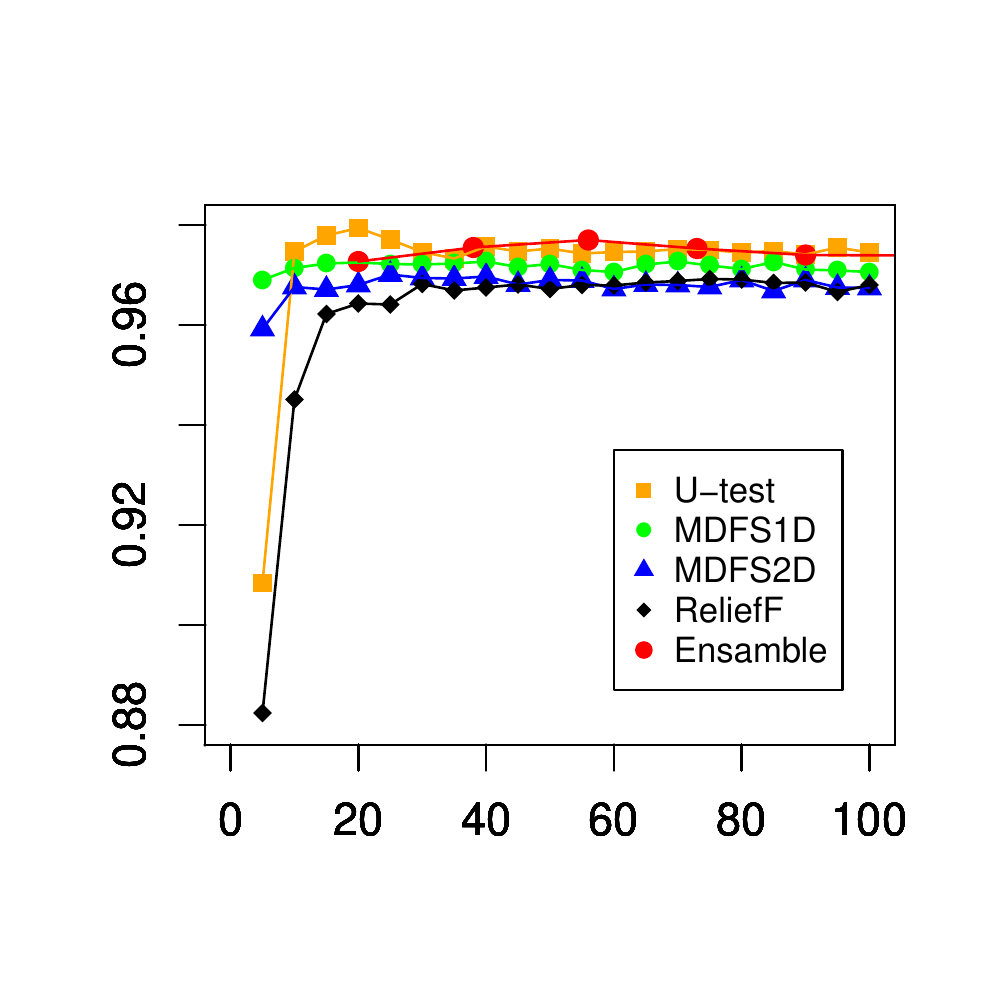} 
  \end{subfigure}%  
  \begin{subfigure}[b]{.45\linewidth}
    \centering
		\caption{HNSC}
    \includegraphics[width=1.15\textwidth,trim ={  .1cm .1cm 0.1cm 1.8cm},clip]{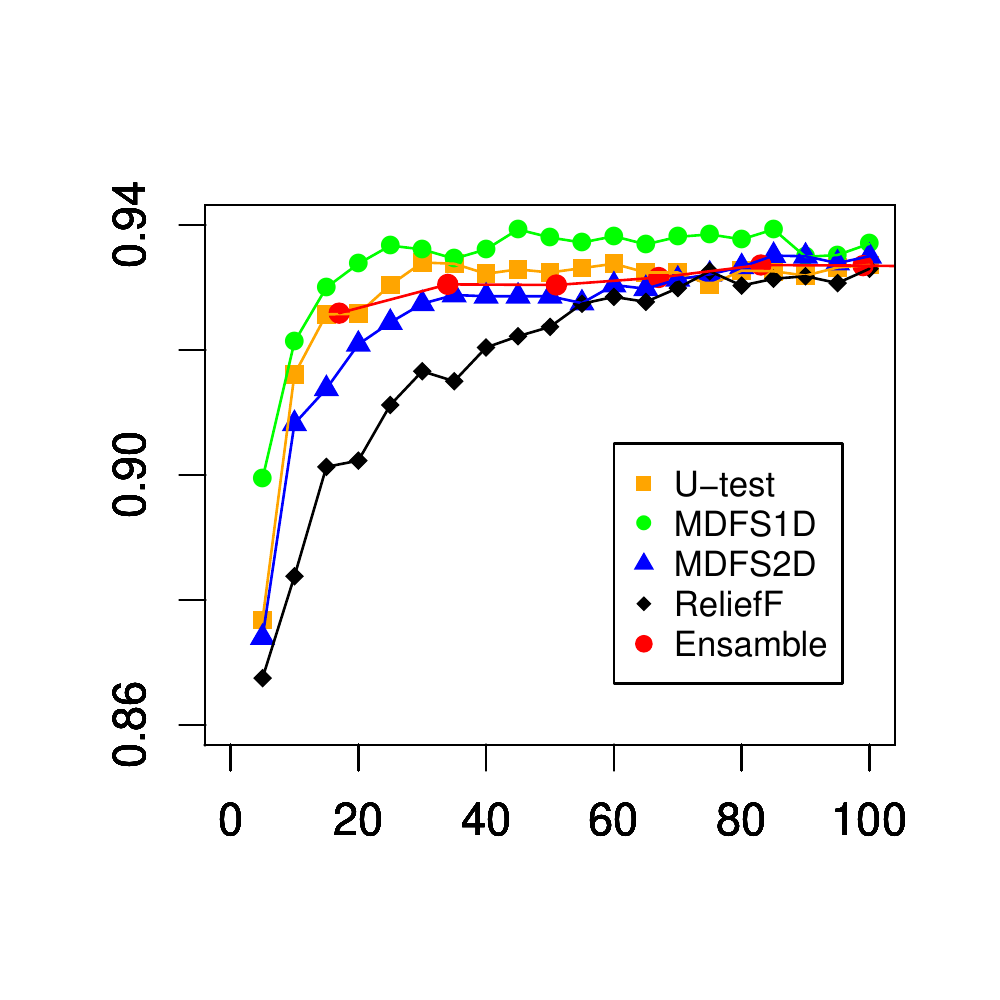} 
  \end{subfigure}\\
  \begin{subfigure}[b]{.45\linewidth}
    \centering
		\caption{KIRC}
    \includegraphics[width=1.15\textwidth,trim ={ .1cm .1cm 0.1cm 1.8cm },clip]{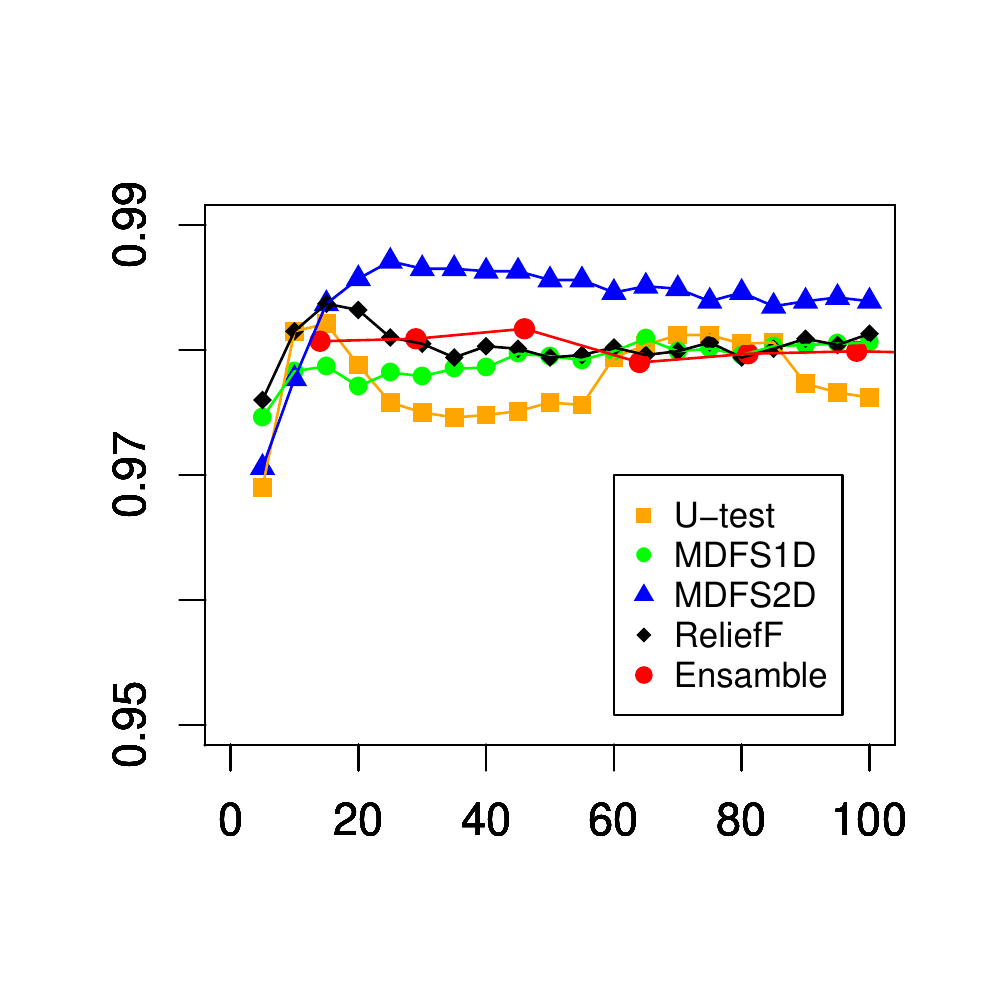} 
  \end{subfigure}
  \begin{subfigure}[b]{.45\linewidth}
    \centering
		\caption{LUAD}
   \includegraphics[width=1.15\textwidth,trim ={  .1cm .1cm 0.1cm 1.8cm},clip]{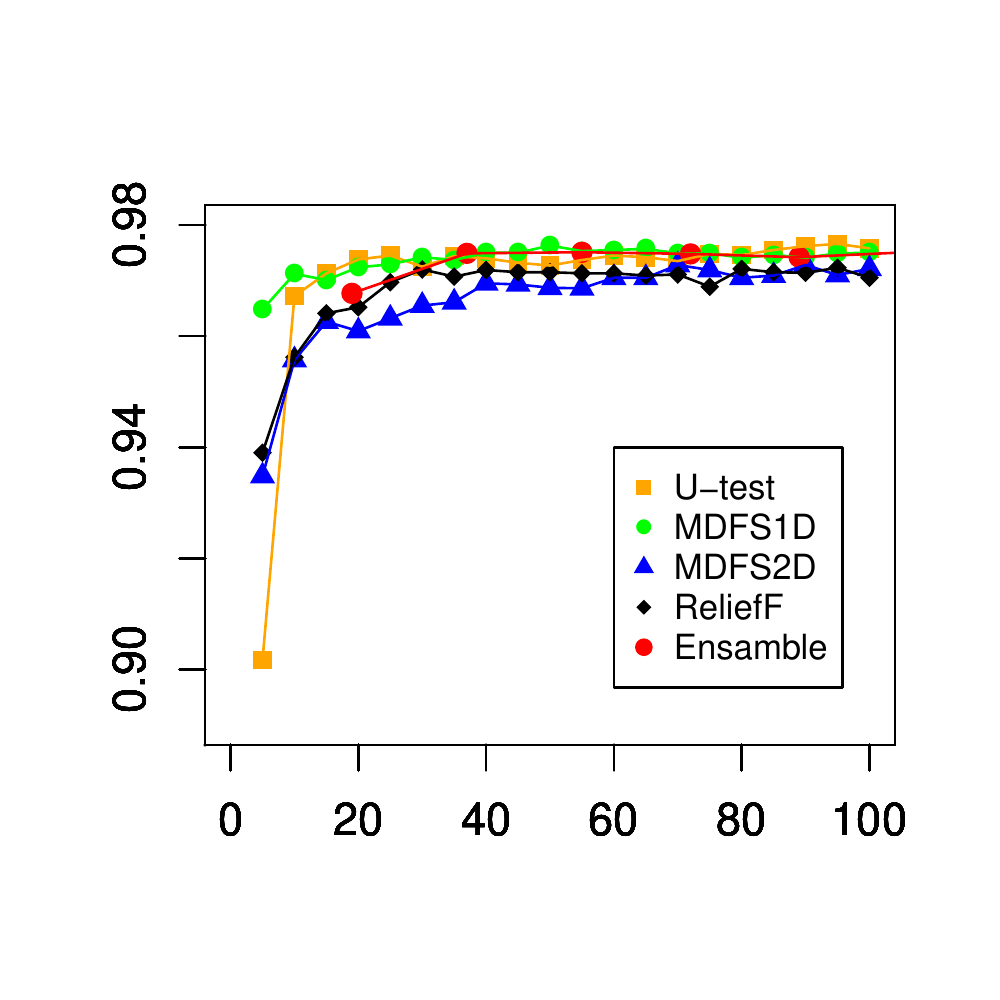} 
  \end{subfigure}
\caption{AUC for models built using top-N variables.}
\label{Figure_AUCvsNtop_Corr075}
\end{figure}

At the next stage of the analysis, the accuracy of models built using top-N features was examined, see Figure~\ref{Figure_AUCvsNtop_Corr075}. 
The number of variables for ensemble model is obtained as the average number of variables in the  union of top-N variables from all FS methods averaged over 150 cross-validated sets.
Generally, the performance of the models is poor for the smallest sizes of variable sets but increases rapidly with increasing number of variables, reaching plateau after roughly 40 variables are included. 
However, there are notable exceptions. 
In particular for BRCA the best model was obtained with 20 variables returned by U-test.  
Even stronger effect was obtained for KIRC data set. 
Here the MDFS-2D feature selection leads to clearly best results at 25 variables, whereas the best results for other filters are obtained with 15 variables. 

\begin{figure}[!t]\flushleft
\centering
  \begin{subfigure}[b]{.4\linewidth}
	  \centering
		\caption{BRCA}
	\includegraphics[width=1.15\textwidth,trim ={ .1cm 0.1cm 0.1cm 0.6cm},clip]{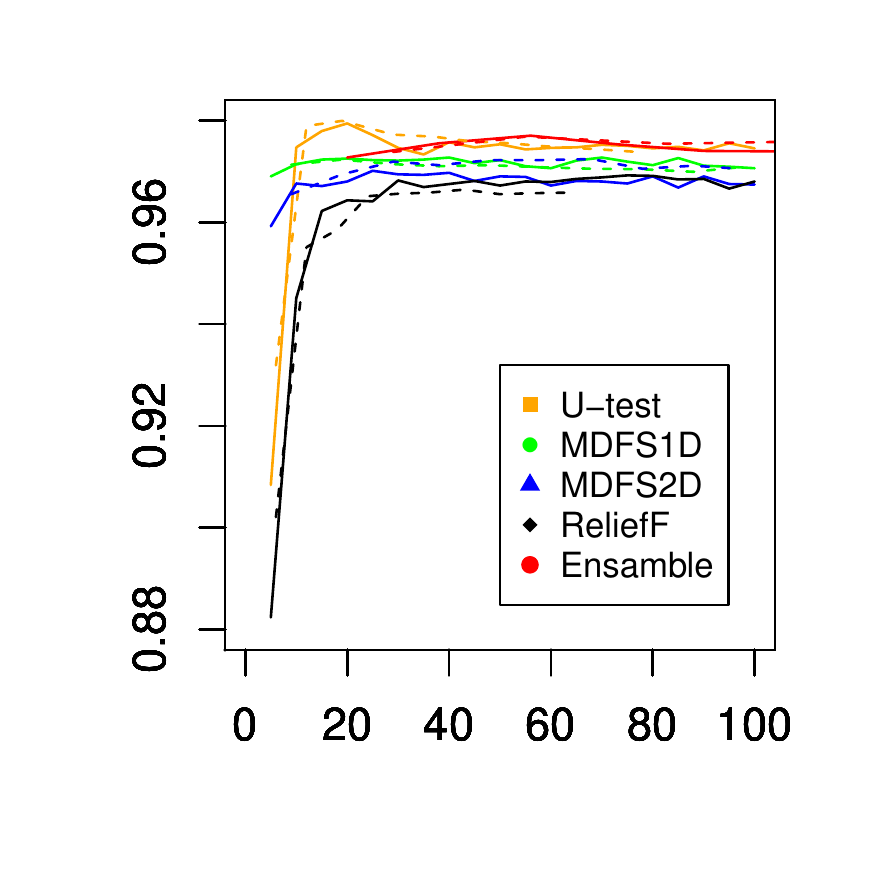} 
  \end{subfigure}%  
  \begin{subfigure}[b]{.4\linewidth}
    \centering
		\caption{HNSC}
    \includegraphics[width=1.15\textwidth,trim ={ .1cm 0.1cm 0.1cm 0.6cm },clip]{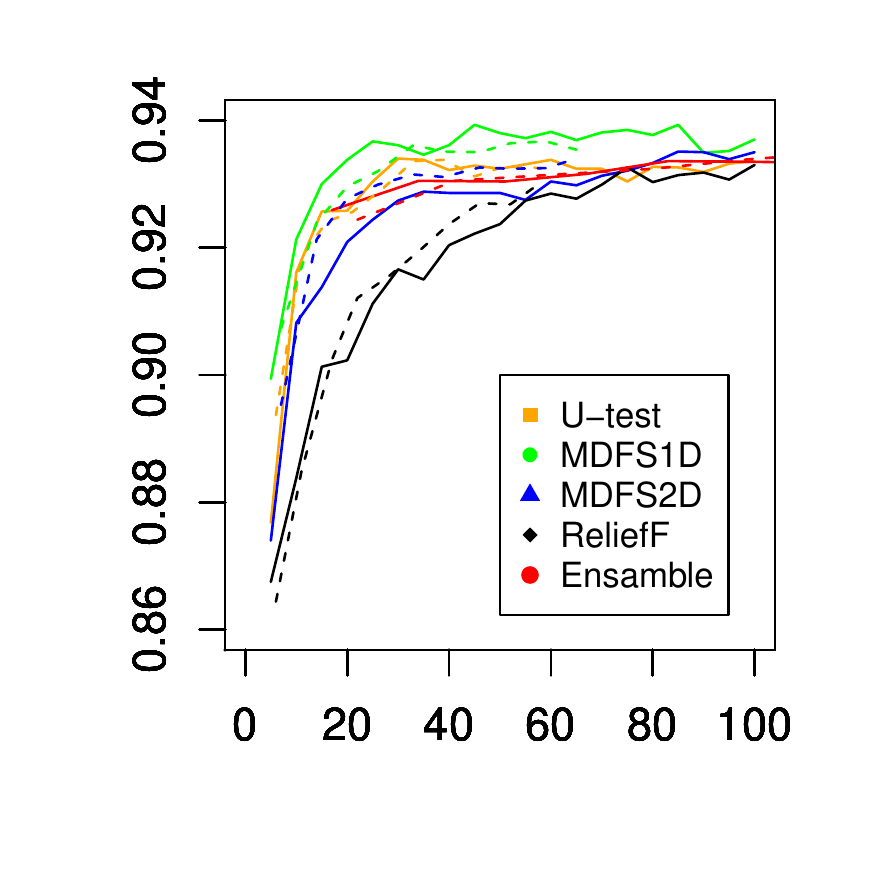} 
  \end{subfigure}\\
  \begin{subfigure}[b]{.4\linewidth}
    \centering
		\caption{KIRC}
    \includegraphics[width=1.15\textwidth,trim ={ .1cm 0.1cm 0.1cm 0.6cm },clip]{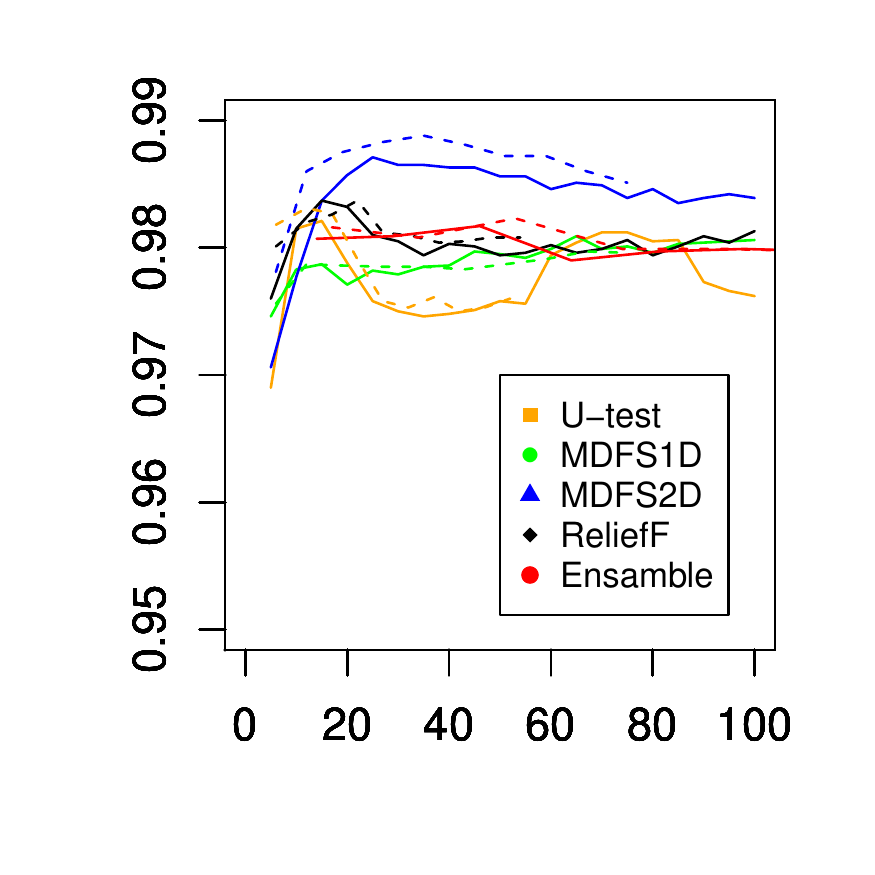}
  \end{subfigure}
  \begin{subfigure}[b]{.4\linewidth}
    \centering
		\caption{LUAD}
   \includegraphics[width=1.15\textwidth,trim ={ .1cm 0.1cm 0.1cm 0.6cm },clip]{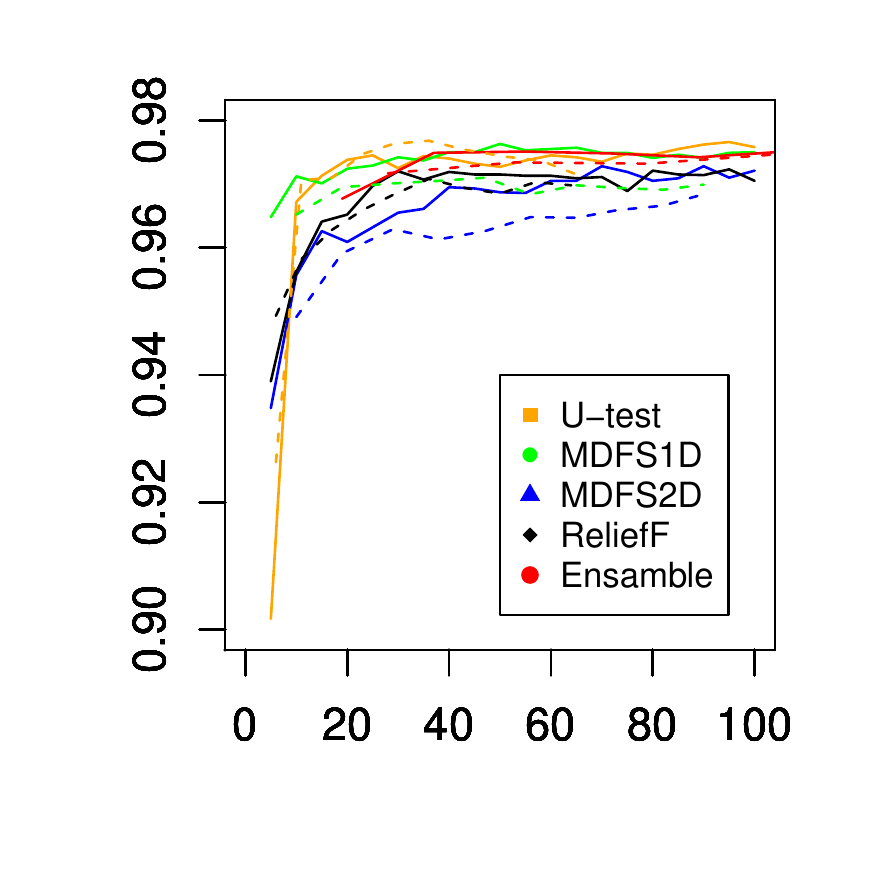}
  \end{subfigure}
\caption{AUC for models built using top-N variables. Solid lines correspond to models built using top-N uncorrelated variables. Dashed lines correspond to models built using top-N uncorrelated variables and variables correlated with them.}

\label{Figure2_AUCvsNtop_Corr075}
\end{figure}

Relative performance of models developed using different FS algorithms vary significantly between data sets. 
For example, the MDFS-2D is clearly the best feature selector for KIRC, and that strongly suggest that non-trivial synergies between variables are present in this data set. 
MDFS-1D is best feature selector for HNSC data set, U-test is best for BRCA, both algorithms are similarly good for LUAD. 

In all cases, the AUC values of models built using variables returned by an ensemble of FS algorithms are comparable to individual models built with a similar number of variables, the AUC curves of the ensemble models (full circle points) are located roughly in the middle of other models as shown in Fig. 3.

Only for BRCA data set for the number of variables larger than 40 the performance of model built on ensemble variables is comparable with the best model for the individual data set, which in this case is a model built using variables from U-test. 

In the next step the effect of adding back redundant variables was examined. 
To this end the RF models were built for the sets of variables consisting of uncorrelated top-N variables and all informative variables highly correlated with them that were previously removed from feature rank list. 
The results are displayed in Figure~\ref{Figure2_AUCvsNtop_Corr075}. 
Clearly adding redundant variables to the main feature set does not improve classification results in most cases. 
An exception are models built using variables obtained with the MDFS-2D method for BRCA, HNSC and KIRC data. 
This effect may arise due to inclusion of correlated variables which interact synergistically with other variables in a slightly different way than those previously included.

\begin{figure}[t]
\centering
  \begin{subfigure}[b]{.45\linewidth}
	  \centering
		\caption{BRCA}
\includegraphics[width=1.15\textwidth,trim ={ .1cm .1cm 0.1cm 2cm},clip]{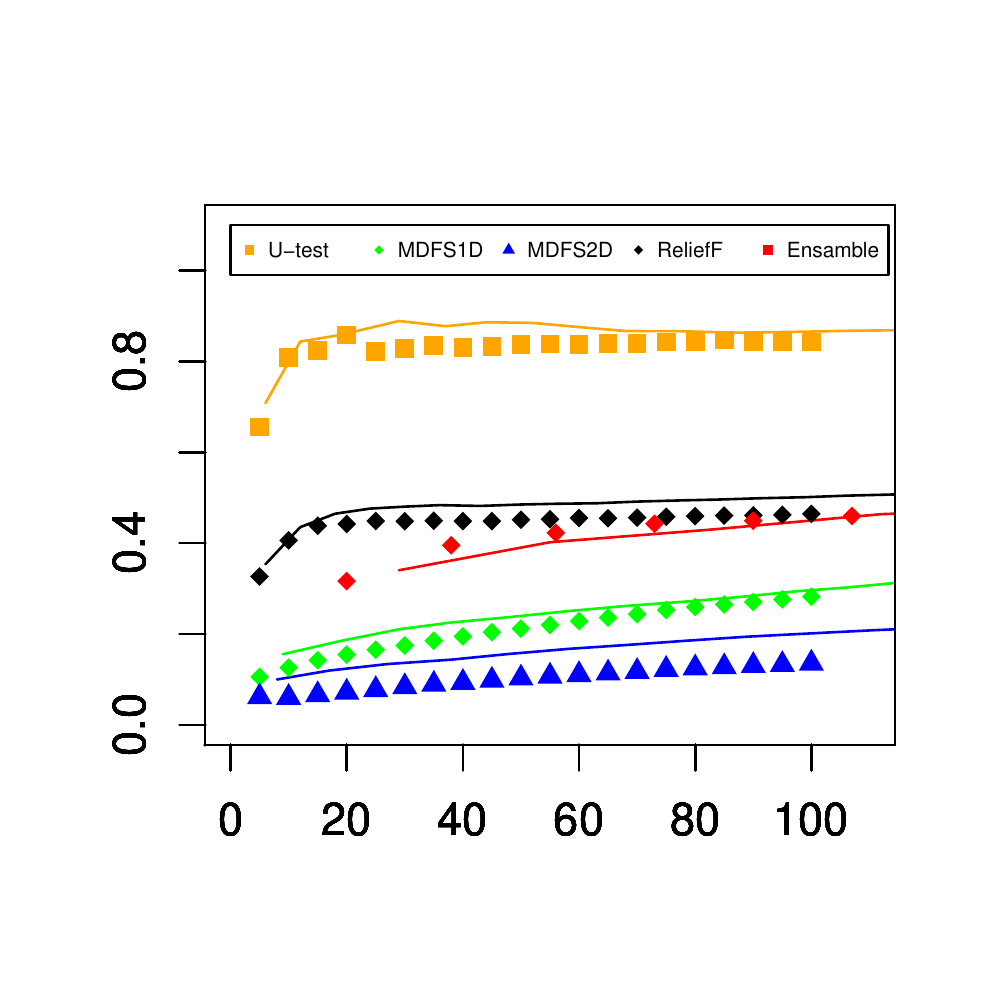} 
  \end{subfigure}%  
  \begin{subfigure}[b]{.45\linewidth}
    \centering
		\caption{HNSC}
    \includegraphics[width=1.15\textwidth,trim ={ .1cm .1cm 0.1cm 2cm },clip]{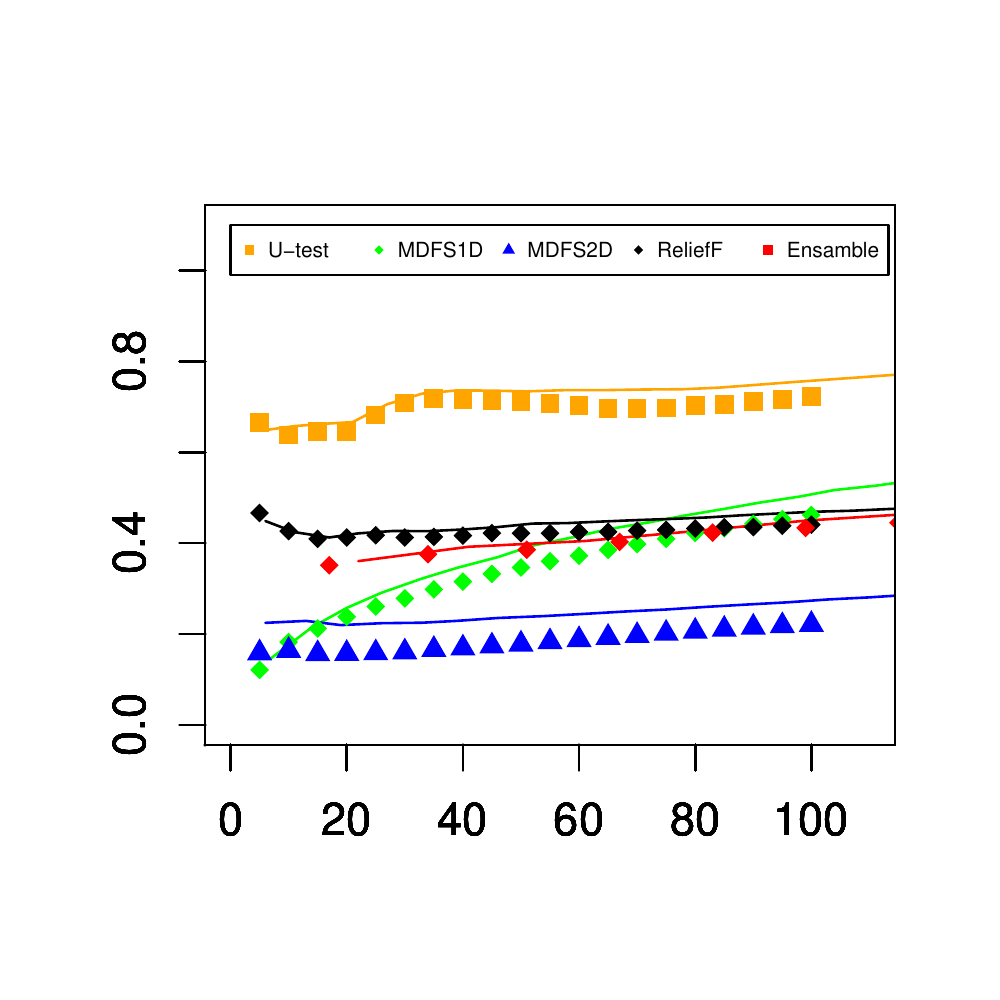} 
  \end{subfigure}\\
  \begin{subfigure}[b]{.45\linewidth}
    \centering
		\caption{KIRC}
    \includegraphics[width=1.15\textwidth,trim ={ .1cm .1cm 0.1cm 2cm },clip]{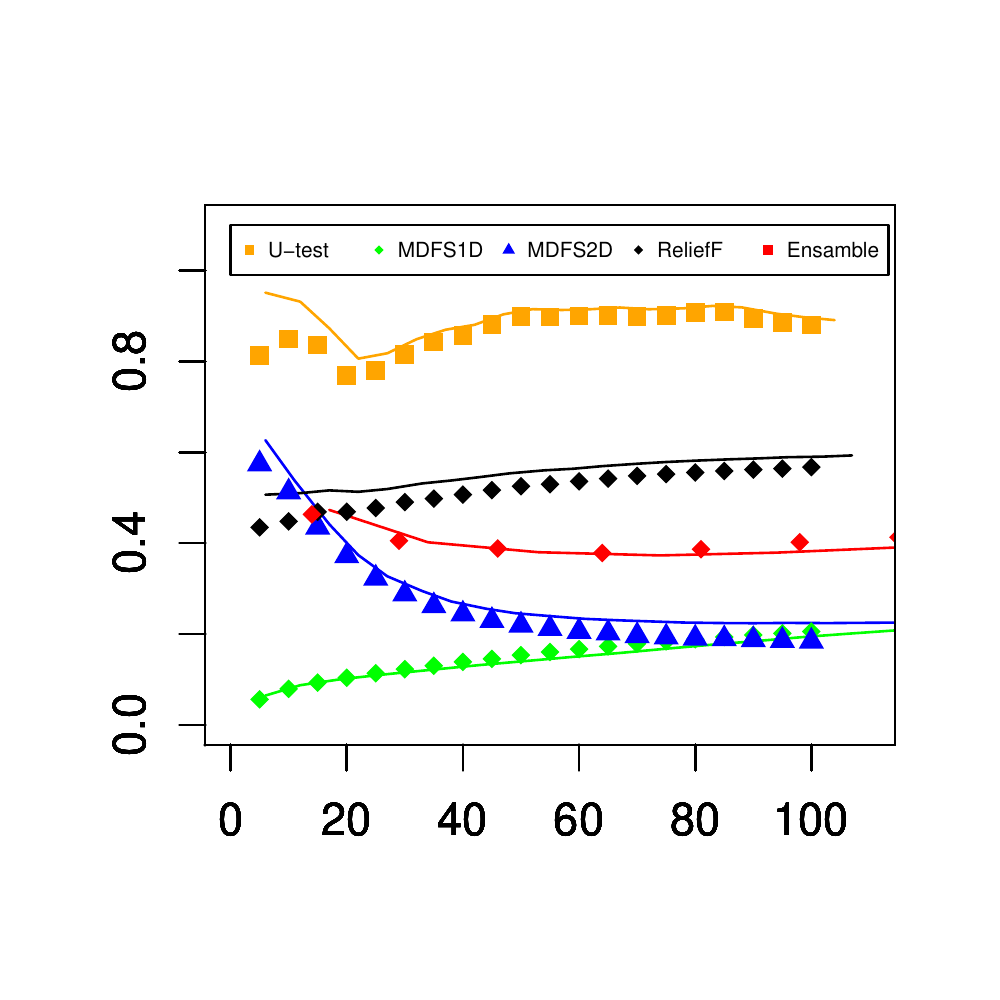}
  \end{subfigure}
  \begin{subfigure}[b]{.45\linewidth}
    \centering
		\caption{LUAD}
   \includegraphics[width=1.15\textwidth,trim ={ .1cm .1cm 0.1cm 2cm },clip]{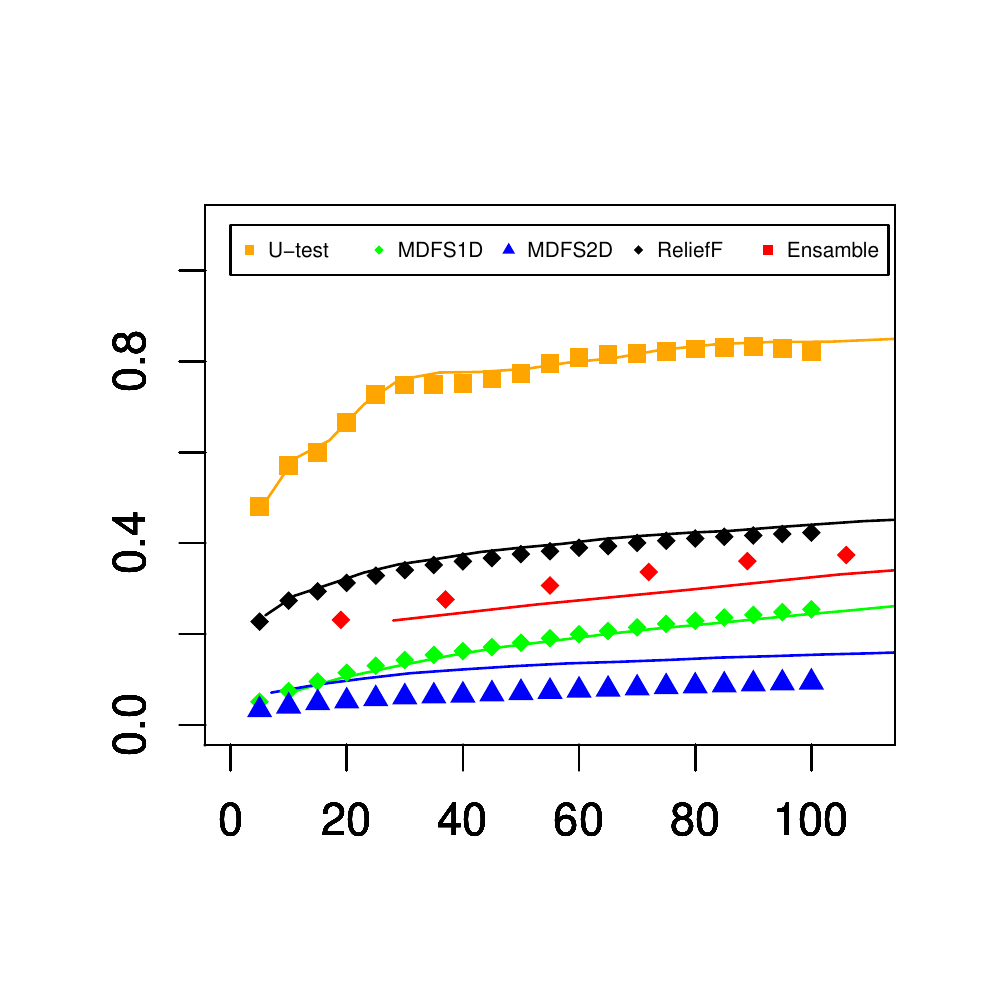}
  \end{subfigure}
\caption{Clock-wise the average similarity (ASM) between 150 feature subsets for top-N variables. Dotted lines correspond to sets consisting from top-N variables. Solid lines correspond to sets consisting from top-N variables and variables highly correlated with them.}
\label{Fig:Similarity_measure_cross_Ntop}
\end{figure}

\subsection{Stability of variable sets}
Often the important property of a feature selection method is stability or robustness of the selected features to perturbations in the data. 
This is particularly important for identification of prognostic or diagnostic markers. 
Therefore the sensitivity of feature selection algorithms to variations in the training sets that arise in the cross-validation were examined. 
The similarity between 150 feature subsets obtained in 150 iterations of cross-validation were measured using the Lustgarten's index ASM, see Figure~\ref{Fig:Similarity_measure_cross_Ntop}.
The highest stability was obtained for variables selected with U-test. 
For this FS method the value of the ASM index varies between 0.7 and 0.8. 
The remaining FS methods are much less stable, with least stable MDFS-2D for which the ASM index is generally below 0.2 and most stable ReliefF for which ASM index varies between 0.3 and 0.5. 
The difference in stability between algorithms is due to the differences in approach used by the algorithm.
The U-test is a deterministic algorithm, for which differences arise exclusively due to variation of sample composition. 
On the other hand all other remaining algorithms rely on randomisation, hence increased variance can be expected. 
In most cases the stability increases with increasing number of variables. 
The notable exception is the MDFS-2D algorithm for KIRC data set, where relatively high stability (ASM$>$0.5) is achieved for smallest size of feature set, and then it rapidly drops with increasing number of variables. 
This result is obtained for the same data set, where models built on feature sets returned by MDFS-2D have highest predictive quality. 
This result strongly suggest existence of a small core of most relevant variables, that must be present in nearly all cases and that strongly contribute to classification. 
This small core is augmented by a diverse group of loosely correlated relevant but redundant variables.  
Finally, in most cases adding redundant variables increases the stability of feature subsets, but the difference is small.  

\subsection{Computational aspects}

The training time does not depend on the type of molecular data, such as microarray gene expression data or DNA copy number data. 
Execution time of the task depends on the size of the dataset, the number of training iterations, the feature selection algorithm, as well as the CPU model or the GPU model.
The most time consuming individual step of the algorithm is feature selection, the model building with Random Forest is relatively quick. 
However, 150 distinct Random Forest were built using the results of the same feature selection step. 
Therefore total time of both components was similar. 
The example execution times for a single iteration of algorithm for KIRC data set are presented in the Table~\ref{tab:exec_time}. 
Among feature selection algorithms used in the study, the ReliefF is by far the most time-consuming. 

\begin{table}[htbp]
\setlength{\tabcolsep}{3.5pt}
%\centering
\caption{Execution times for a single iteration of the algorithm for the KIRC data set. Computations performed on a CPU Intel Xeon Processor E5-2650v2. The MDFS-2D algorithm was executed using a GPU-accelerated version on NVIDIA Tesla K80 co-processor.} 
\label{tab:exec_time}
\begin{tabular}{|l | l | l | l | l | l | l | l |}
U-test     &  MDFS-1D &  MDFS-2D & ReliefF & Ensemble & RF $\times$1 & RF $\times$ 100 & Total \\
\hline
00m:37s    &  00m:04s &  00m:03s & 05m:41s & 05m:54s & 00m:03s      &  05m:19s  &  10m:13s \\
 \end{tabular}
\end{table}

The single run of the algorithm involved calling four FS methods, removing correlated features, producing a ranking of the features, and calling RF classiﬁcation algorithm 150 times (5 feature sets × 20 sizes of feature set). The algorithm was executed 150 times, computations for one data set took about 25 hours of CPU time.

\section{Conclusions}

The current study demonstrates that relying on a single FS algorithm is not optimal.
Different FS algorithms are best suited for identification of the most relevant feature sets in various data sets. 
Combining variables from multiple FS algorithms into a single feature set does not improve the performance in comparison with an equally numerous feature set generated by the individual algorithm that is best suited for the particular data set.  
On the other hand, application of multiple algorithms increases the chances of identifying the best FS algorithm for the  problem under scrutiny. 
In particular, application of the FS algorithms that can detect synergies in the data can significantly improve the quality of machine learning models. 

Interestingly, the stability of a FS algorithm is not required for building a highly predictive machine learning models. 
This is possible, since biological systems often contain multiple informative variables. 
Therefore, useful models can be obtained using very diverse combinations of predictive variables. 

\section*{Notes}
\subsection*{Acknowledgements}
This work was supported by the National Science Centre, Poland in frame of grant Miniatura 2 No. 2018/02/X/ST6/02571

\bibliography{bib_Polewko}

\end{document}